\documentclass[twocolumn,showpacs,aps,prl,superscriptaddress,floatfix,letter]{revtex4}

\usepackage{epsfig}
\usepackage{graphics}
\usepackage{graphicx}
\usepackage{epic}
\usepackage{floatflt}
\usepackage{dcolumn}
\usepackage{amsmath}
\usepackage{graphpap}
\usepackage{feynmf}
\usepackage{multirow}
\usepackage{color}

\input babarsym
\def\fish {\ensuremath{\cal F}\xspace}

\long\def\inst#1{\par\nobreak\kern 4pt\nobreak
    {\it #1}\par\vskip 10pt plus 3pt minus 3pt}

\newcommand{\BaBarYear}       {10}
\newcommand{\BaBarNumber}     {001}
\newcommand{\SLACPubNumber} {14086}
 \newcommand{\BaBarType}      {PUB}  

\begin{document}

\par\vskip 7cm

\begin{flushleft}
\babar-\BaBarType-\BaBarYear/\BaBarNumber \\
SLAC-PUB-\SLACPubNumber\\
\end{flushleft}
\title{\boldmath Search for $\Bp \to\ \Dp K^0$ and $\Bp \to\ \Dp \Kstarz$ decays}
%
\author{P.~del~Amo~Sanchez}
\author{J.~P.~Lees}
\author{V.~Poireau}
\author{E.~Prencipe}
\author{V.~Tisserand}
\affiliation{Laboratoire d'Annecy-le-Vieux de Physique des Particules (LAPP), Universit\'e de Savoie, CNRS/IN2P3,  F-74941 Annecy-Le-Vieux, France}
\author{J.~Garra~Tico}
\author{E.~Grauges}
\affiliation{Universitat de Barcelona, Facultat de Fisica, Departament ECM, E-08028 Barcelona, Spain }
\author{M.~Martinelli$^{ab}$}
\author{A.~Palano$^{ab}$ }
\author{M.~Pappagallo$^{ab}$ }
\affiliation{INFN Sezione di Bari$^{a}$; Dipartimento di Fisica, Universit\`a di Bari$^{b}$, I-70126 Bari, Italy }
% ...

\begin{abstract}

\pacs{13.25.Hw, 14.40.Nd} We report a search for the rare decays
$\Bp \to \Dp K^0$ and $\Bp \to \Dp \Kstarz$ in an event sample of
approximately $465$ million \BB pairs collected with the \babar\
detector at the PEP-II asymmetric-energy \epem collider at SLAC. We
find no significant evidence for either mode and we set 90\%
probability upper limits on the branching fractions of $\BR(\Bp \to\
\Dp K^0) < 2.9\times 10^{-6}$ and $\BR(\Bp \to\ \Dp \Kstarz) <
3.0\times 10^{-6}$.
\end{abstract}

\maketitle

\section{Introduction}

Charged \B meson decays in which neither constituent quark appears
in the final state, such as $\Bp \to \Dp K^{(*)0}$, are expected to
be dominated by weak annihilation diagrams with the $\bbar u$ pair
annihilating into a \Wp boson. Such processes therefore can provide
insight into the internal dynamics of \B mesons, in particular the
overlap between the $b$ and the $u$ quark wave functions.
Annihilation amplitudes cannot be evaluated with the commonly-used
factorization approach~\cite{cite:buras}. As a consequence, there
are no reliable estimates for the corresponding decay rates.
Annihilation amplitudes are expected to be proportional to $f_B/m_B$
where $m_B$ is the mass of the $B$ meson and $f_B$ is the
pseudoscalar \B meson decay constant. The quantity $f_B$ represents
the probability amplitude for the two quark wave functions to
overlap. Numerically, $f_B/m_B$ is approximately equal to
$\lambda^2$, where $\lambda$ is the sine of the Cabibbo
angle~\cite{cite:gronau-anni, cite:buras}. In addition, these
amplitudes are also suppressed by the CKM factor $\Vub
\sim\lambda^3$. So far, there has been no observation of a hadronic
\B meson decay that proceeds purely through weak annihilation
diagrams, although evidence for the leptonic decay $\B\to\tau\nu$
has been found~\cite{cite:taunu}. In theoretical calculations of
nonleptonic decays, the assumption is often made that these
amplitudes may be neglected.

\begin{figure}[htb]
\begin{center}
\begin{fmffile}{graph1}
\vskip 30pt
\begin{fmfgraph*}(100,50)
\fmfleft{bbar,u} \fmf{fermion, tension=1/2,label=$u$,
lab.sid=left}{u,Zee} \fmf{fermion, tension=1/2,label=$\bar b$,
lab.sid=left}{Zee,bbar} \fmf{boson,label=$W^{+}$,
tension=7/8}{Zee,Zff} \fmf{fermion,label=$c$, lab.sid=left,
tension=1/4}{Zff,c} \fmf{fermion,label=$\bar s$, lab.sid=left,
tension=1/4}{sbar,Zff} \fmf{phantom}{Zff,gammab} \fmf{phantom,
tension=3/8}{gammab,gammaf} \fmf{fermion, tension=1/4,label=$d$,
lab.sid=right}{gammaf,d} \fmf{fermion, tension=1/4,label=$\bar
d$,lab.sid=right}{dbar,gammaf} \fmfright{sbar,d,dbar,c}
\fmfdot{Zee,Zff,gammaf} \fmffreeze \fmf{phantom,label=$B^{+}$,
lab.sid=left,lab.dist=.1w}{bbar,u} \fmf{phantom,label=$\Dp$,
lab.sid=left,lab.dist=.1w}{c,dbar} \fmf{phantom,label=$K^{(*)0}$,
lab.sid=left,lab.dist=.1w}{d,sbar}
\end{fmfgraph*}\vskip 30pt
\begin{fmfgraph*}(100,50)
\fmflabel{}{bbar}
\fmflabel{}{u1}
\fmflabel{}{c}
\fmflabel{}{u2}
\fmflabel{}{ubar}
\fmflabel{}{sbar}
\fmfleft{v1,v2,ph1,ph2}
\fmf{fermion,
tension=0.25,label=$\bar{u}$,lab.sid=right,lab.dist=.04w}{ubar,Zbb}
\fmf{fermion, tension=0.25,label=$\bar b$, lab.sid=right}{Zbb,bbar}
\fmf{fermion, tension=0.25,label=$u$, lab.sid=right}{u1,Abb,u2}
\fmf{phantom, tension=0.25,label=$\pi^0$, lab.sid=right}{u2,ubar}
\fmf{phantom, tension=0.25,label=$B^+$, lab.sid=left}{u1,bbar}
\fmf{phantom, tension=10}{v1,u1}
\fmf{phantom, tension=1}{v2,bbar}
\fmf{phantom, tension=10}{o1,u2}
\fmf{phantom, tension=1}{o2,ubar}
\fmfright{o1,o2,sbar,c}
\fmffreeze
\fmf{boson, tension=2, label=$W^{+}$, lab.sid=left}{Zbb,Zff}
\fmf{fermion, label=$c$, lab.sid=left}{Zff,c}
\fmf{fermion, label=$\bar s$, lab.sid=left}{sbar,Zff}
\fmf{phantom, label=$D_s$, lab.sid=left}{c,sbar}
\fmfdot{Zbb,Zff}
\end{fmfgraph*}\hskip 30pt
\begin{fmfgraph*}(100,50)
\fmflabel{$B^+$}{B}
\fmflabel{$D^{+}$}{D}
\fmflabel{$K^{(*)0}$}{K}
\fmfleft{B}
\fmfright{D,K}
\fmf{fermion, tension=2.5}{B,LL}
\fmf{fermion,label=$\Ds$, lab.sid=right}{LL,LUP}
\fmf{fermion, tension=2.5}{LUP,D}
\fmf{fermion,label=$\pi^0$, lab.sid=left}{LL,LDOWN}
\fmf{fermion,label=$K^{*0}$}{LUP,LDOWN}
\fmf{fermion, tension=2.5}{LDOWN,K}
\fmfdot{LL,LUP,LDOWN}
\end{fmfgraph*}
\end{fmffile}
\vskip 10pt \caption{\label{diagr}Annihilation diagram for the decay
$B^+{\to}\Dp K^{(*)0}$ (top). Tree diagram (bottom left) for the
decay $B^+ \to D_s^+ \pi^0$ and hadron-level diagram (bottom right)
for the rescattering contribution to $B^+{\to}D^{+} K^{(*)0}$ via
$B^+{\to}\Ds \pi^0$.}
\end{center}
\end{figure}
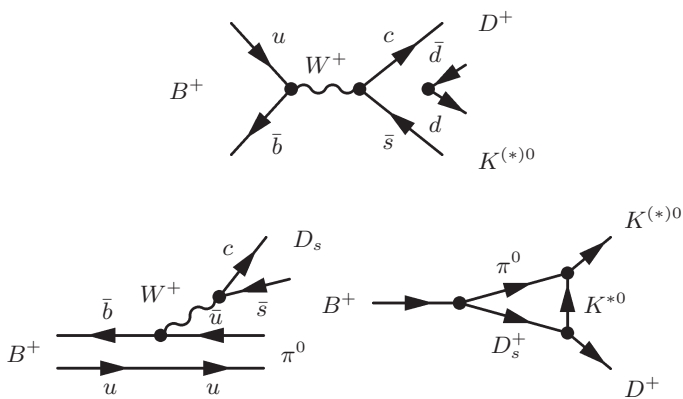

Some studies indicate that the branching fractions of
weak-annihilation processes could be enhanced by so-called
rescattering effects, in which long-range strong interactions
between \B decay products, rather than the decay amplitudes, lead to
the final state of interest~\cite{cite:gronau-anni}. Figure
\ref{diagr} shows the Feynman diagram for the decays ${\ensuremath
{\Bp{\to}\Dp K^{(*)0}}}$ and ${\ensuremath
{B^+{\to}D_s^{+}\pi^0}}$~\cite{cite:ch}, and the hadron-level
diagram for the rescattering of $\Ds\piz$ into $\Dp K^{(*)0}$.
Significant rescattering could thus mimic a large weak annihilation
amplitude. It has been argued~\cite{cite:gronau-anni} that
rescattering effects might be suppressed by only $\lambda^4$,
compared to $\lambda^5$ for the weak annihilation amplitudes,
rendering the $\Bp \to \Dp K^{(*)0}$ decay rate due to rescattering
comparable to the isospin-related color-suppressed $\Bz \to \Dz
K^{(*)0}$ decay rate of  approximately $5 \times 10^{-6}$.

$\Bp\to\Dp K^{(*)0}$ decays are also of interest because their decay
rates can be used to constrain the annihilation amplitudes in
phenomenological fits~\cite{cite:buras,cite:us}. This allows the
translation of the measurements of the ${\ensuremath
{B^+{\to}D^{0}K^{(*)+}}}$ amplitudes into estimations of the \Vub\
suppressed amplitudes ${\ensuremath
{B^0{\to}D^{0}K^{(*)0}}}$~\cite{cite:us, cite:d0k0}. None of the
modes studied here has been observed so far, and a 90\% confidence
level upper limit on the branching fraction ${\ensuremath {{\cal
B}(\Bp{\to}\Dp K^0})} < 5\times10^{-6}$ has been established by
\babar~\cite{cite:polci}.  No study of $\Bp\to\Dp\Kstarz$ has
previously been published.

The results presented here are obtained with 426~fb$^{-1}$ of data
collected at the $\FourS$ resonance with the \babar\ detector at the
\pep2\ asymmetric \epem collider~\cite{cite:pep2} corresponding to
$465 \times 10^6$ \BB pairs ($N_{\BB}$). An additional 44.4~\invfb
of data (``off-resonance'') collected at a center-of-mass (CM)
energy 40~\mev below the $\FourS$ resonance is used to study
backgrounds  from $e^+ e^- \to \qqbar$ ($q=u,\,d,\,s,$ or $c$)
processes, which we refer to as continuum events.

The \babar\ detector is described in detail elsewhere
\cite{cite:det}. Charged-particle tracking is provided by a five
layer silicon vertex tracker (SVT) and a 40 layer drift chamber
(DCH). In addition to providing precise position information for
tracking, the SVT and DCH measure the specific ionization, which is
used for particle identification of low-momentum charged particles.
At higher momenta ($p > 0.7\gevc$) pions and kaons are identified by
Cherenkov radiation detected in a ring-imaging device (DIRC). The
position and energy of photons are measured with an electromagnetic
calorimeter (EMC) consisting of 6580 thallium-doped CsI crystals.
These systems are mounted inside a 1.5~T solenoidal super-conducting
magnet. Muons are identified by the instrumented magnetic-flux
return, which is located outside the magnet.

\section{Event Reconstruction and Selection}

The event selection criteria are determined using Monte Carlo (MC)
simulations of $\epem\to\FourS\to\BB$ (``\BB'' in the following) and
continuum events, and the off-resonance data. The selection criteria
are optimized by maximizing the quantity $S/\sqrt{S+B}$, where $S$
and $B$ are the expected numbers of signal and background events,
respectively. We assume the signal branching fraction to be $5\times
10^{-6}$ in the optimization procedure.

The charged particle candidates are required to have transverse
momenta above 100~\mevc and at least twelve hits in the DCH.

Candidate \Dp mesons are reconstructed in the $\Dp \to
K^-\pi^+\pi^+$ ($K\pi\pi$ in the following), $\Dp \to \KS\pi^+$
($\KS\pi$), $\Dp \to K^-\pi^+\pi^+ \pi^0$ ($K\pi\pi\pi^0$) and $\Dp
\to \KS\pi^+ \pi^0$ ($\KS\pi\pi^0$) modes for the $\Bp\to\Dp\Kz$
decay channel ($DK$). Only the first two modes are used for the $\Bp
\to\Dp\Kstarz$ decay channel ($DK^*$ in the following) since we find
that including the $K\pi\pi\pi^0$ and $\KS\pi\pi^0$ modes in this
channel does not appreciably improve the sensitivity of the
analysis.

The \Dp candidates are reconstructed by combining kaons (either
charged or neutral depending on the channel) and the appropriate
number of pions. The charged kaons used to reconstruct the \Dp and
$K^{*0}$ candidates are required to satisfy kaon identification
criteria obtained using a likelihood technique based on the opening
angle of the Cherenkov light measured in the DIRC and the ionization
energy loss measured in the SVT and DCH. These criteria are
typically 85\% efficient, depending on the momentum and polar angle,
with misidentification rates at the 2\% level. Kaons and pions from
$D$ decays are required to have momenta in the laboratory frame
greater than 200~$\mevc$ and 150~$\mevc$, respectively. The
reconstructed \Dp candidates are required to satisfy the invariant
mass ($M_{\rm D}$) selection criteria given in Table~\ref{tab:cuts}.

The \KS candidates are reconstructed from pairs of
oppositely-charged pions with invariant mass within 5--7 \mevcc\ of
the nominal \KS mass \cite{cite:PDG}. This mass cut corresponds to
2--2.8 standard deviations of the experimental resolution and varies
slightly among channels due to the different amounts of background
per channel. For the prompt \KS candidates from the $\Bp\to\Dp\KS$
decay, we require $1-\cos \alpha_{\KS}(\B^+)<10^{-8}$, where
$\alpha_{\KS}(\B^+)$ is the angle between the momentum vector of the
\KS candidate and the vector connecting the \Bp and \KS decay
vertices. For \KS daughters of a \Dp decay, we require $1-\cos
\alpha_{\KS}(\Dp)<10^{-6}$, where $\alpha_{\KS}(\Dp)$ is defined in
a similar way.

The $\pi^0$ candidates are reconstructed from pairs of photon
candidates each with an energy greater than 70~\mev, and a lateral
shower profile in the EMC consistent with a single electromagnetic
deposit. These pairs must have a total energy greater than 200\mev,
a CM momentum greater than 400~\mevc, and an invariant mass within
10\mevcc (for the $K\pi\pi\pi^0$ mode) or 12\mevcc (for the
$\KS\pi\pi^0$ mode) of the nominal $\pi^0$ mass~\cite{cite:PDG}.

The $\Kstarz$ candidates are reconstructed in the decay channel
$\Kstarz \to \Kp\pim$. These charged tracks are constrained to
originate from a common vertex.  The reconstructed invariant mass,
whose width is dominated by the \Kstarz natural width, is required
to lie within $40~\mevcc$ of the nominal \Kstarz
mass~\cite{cite:PDG}. We define $\theta_{\rm H}$ as the angle
between the direction of flight of the charged $K$ and the direction
of flight of the $B$ in the $\Kstarz$ rest frame. The probability
distribution of $\cos\theta_{\rm H}$ is proportional to
$\cos^2\theta_{\rm H}$ for longitudinally polarized $K^{*0}$ mesons
from $\B\to D K^{*0}$ decays, due to angular momentum conservation,
and is approximately flat for fake (random combinations of tracks)
or unpolarized background $K^{*0}$ candidates. To suppress fake and
background $K^{*0}$ candidates we require $|\cos\theta_{\rm
H}|>0.5$.

The $B^+$ candidates are reconstructed by combining one \Dp and one
$\KS$ or $K^{*0}$ candidate, constraining them to originate from a
common vertex. The probability distribution of the cosine of the $B$
polar angle with respect to the beam axis in the CM frame,
$\cos\theta_{B}$, is expected to be proportional to
$1-\cos^2\theta_{B}$. Selection criteria on $|\cos\theta_{B}|$ are
channel dependent and are summarized in Table \ref{tab:cuts}.

We measure two almost independent kinematic variables: the
beam-energy substituted mass
$\mes\equiv\sqrt{(E^{*2}_{0}/2+\vec{p_0}\cdot\vec{p_{\B}})^2/E^{2}_{0}-{p_{\B}}^2}$,
and the energy difference $\DeltaE \equiv E^{*}_B-E^*_{0}/2$, where
$E$ and $p$ are energy and momentum, the subscripts $B$ and $0$
refer to the candidate $B$ and to the $e^+e^-$ system, respectively,
and the asterisk denotes a calculation made in the CM frame. Signal
events are expected to peak at the $B$ meson mass for \mes\ and at
zero for $\Delta E$. Channel-dependent selection criteria on
$|\Delta E|$ are given in Table~\ref{tab:cuts}. We retain candidates
with \mes\ in the range $[5.20,5.29]$ \gevcc for subsequent
analysis.

\begin{table*}[htb]
\begin{center}
\caption{\label{tab:cuts}Main selection criteria used to distinguish
between signal and background events. $M_{D,\,{\rm PDG}}$ is the
nominal mass of the \Dp meson~\cite{cite:PDG}.}
\begin{tabular}{lcccc@{\hspace{8mm}}cc}
\hline\hline\small
\multirow{2}{*}{Selection criteria} &\multicolumn{4}{c}{$B^+\to \Dp K^0$}&\multicolumn{2}{c}{$B^+\to \Dp K^{*0}$}\\
\cline{2-7} &${K\pi\pi}$ & ${K\pi\pi\piz}$ & ${\KS\pi}$ & ${\KS\pi\piz}$ &${K\pi\pi}$ & ${\KS\pi}$\\
\hline
$|M_{D,\,{\rm PDG}}|$ (\mevcc) &$<$12 ($\simeq$ 1.9$\sigma$)&$<$18 ($\simeq$ 1.5$\sigma$) &$<$14 ($\simeq$ 1.6$\sigma$) &$<$22 ($\simeq$ 1.6$\sigma$) &$<$10 ($\simeq$ 1.6$\sigma$) &$<$10 ($\simeq$ 1.8$\sigma$) \\
$|\cos\theta_{B}|$ &$<$0.76 &$<$0.77 &$<$0.87 &$<$0.85 &$<$0.82 &$<$0.84 \\
$|\Delta E |$ (MeV) &$<$20 ($\simeq$ 1.3$\sigma$) &$<$23 ($\simeq$ 1.5$\sigma$) &$<$25 ($\simeq$ 1.5$\sigma$) &$<$24 ($\simeq$ 1.5$\sigma$) &$<$19 ($\simeq$ 1.3$\sigma$) &$<$19 MeV ($\simeq$ 1.3$\sigma$) \\
\hline\hline
\end{tabular}
\end{center}
\end{table*}

\begin{table*}[htb]
\begin{center}
\caption{\label{tab:eff}Reconstruction efficiencies and expected
numbers of events in the fit and signal region assuming $\BR(B^+\to
\Dp K^0)=\BR(B^+\to \Dp K^{*0})=5\times10^{-6}$.}
\begin{tabular}{lcr@{ $\pm$ }lr@{ $\pm$ }lr@{ $\pm$ }lr@{ $\pm$ }l@{\hspace{8mm}}r@{ $\pm$ }lr@{ $\pm$ }l}
\hline\hline
\multirow{2}{*}{} &\multirow{2}{*}{region}&\multicolumn{8}{c}{$B^+\to \Dp K^0$} &\multicolumn{4}{c}{$B^+\to \Dp K^{*0}$}\\
\cline{3-14} & &\multicolumn{2}{c}{${K\pi\pi}$} &\multicolumn{2}{c}{${K\pi\pi\pi^0}$} &\multicolumn{2}{c}{${\KS\pi}$} &\multicolumn{2}{c}{${\KS\pi\pi^0}$} &\multicolumn{2}{c}{${K\pi\pi}$} &\multicolumn{2}{c}{${\KS\pi}$}\\
\hline
\multirow{2}{*}{Signal efficiency} &fit & \multicolumn{2}{c}{18.4\%} & \multicolumn{2}{c}{5.2\%} & \multicolumn{2}{c}{21.3\%} & \multicolumn{2}{c}{6.2\%} & \multicolumn{2}{c}{10.6\%} & \multicolumn{2}{c}{10.5\%} \\
 &signal & \multicolumn{2}{c}{12.4\%} & \multicolumn{2}{c}{3.8\%} & \multicolumn{2}{c}{14.7\%} & \multicolumn{2}{c}{4.9\%} & \multicolumn{2}{c}{7.6\%} & \multicolumn{2}{c}{7.4\%}
 \\\hline
\multirow{2}{*}{Signal} &fit & 14.1 & 0.2 & 2.5&0.1 & 1.81&0.03 & 2.4&0.1 & 15.8 & 0.3 & 1.70&0.04 \\
 &signal & 9.6 & 0.2 & 1.8&0.1 & 1.21&0.03 & 1.9&0.1 & 11.3&0.3 & 1.20&0.03 \\\hline
\multirow{2}{*}{Combinatorial \BB background} &fit & 67 & 4 & 157&4 & 12&2 & 36&3 & 400&10 & 42.8&4 \\
 &signal & 7 & 2 & 20&2 & 3&1 & 8&2 & 30&2 & 6.4&1 \\\hline
\multirow{2}{*}{Peaking \BB background} &fit & 2.0 & 0.2 & 3.3&0.4 & 1.1&0.2 & 1.8&0.5 & 26&2 & 2.4&0.3 \\
 &signal & 0.3 & 0.1 & 1.0&0.2 & 0.3&0.1 & 0.6&0.2 & 5.4&1 & 0.7&0.2 \\\hline
\multirow{2}{*}{Continuum background} &fit & 2840 & 40 & 4860&50 & 640&20 & 1600&30 & 6100&50 & 630&20 \\
 &signal & 63 & 6 & 104&8 & 12&3 & 45&5 & 129&8 & 13&3 \\
\hline\hline
\end{tabular}
\end{center}
\end{table*}

In less than 1\% of the cases, multiple \Bp candidates are present
in the same event, and in those cases we choose the one with the
reconstructed \Dp mass closest to the nominal mass
value~\cite{cite:PDG}. If more than one \Bp candidate shares the
same \Dp candidate, then we choose the \Bp candidate with \DeltaE
closest to zero.

\begin{table*}[htb]
\footnotesize
\begin{center}
\caption{Expected errors on the branching fractions from toy MC
studies depending on the branching fractions generated. The combined
errors are obtained as results of likelihood combination per each
toy (see text for details). All the numbers are given in units of
$10^{-6}$.\label{tab:errors}}
\begin{tabular}{lcc}
\hline\hline
\multirow{2}{1in}{Decay mode} &$\BR=5$ & \BR=0 \\
\cline{2-3} & Mean error [95\% range] & {Mean error [95\% range]} \\

\hline \multicolumn{3}{l}{$B^+ \to \Dp K^0$ }\\\hline

\multirow{2}{1in}{~~${K\pi\pi}$} &${+3.3~~ [2.7, 4.0]}$ &${+2.8~~ [2.2, 3.6]}$ \\
 &${-3.0 ~~[2.2, 3.6]}$ &${-2.4~~ [1.6, 3.2]}$ \\\hline
\multirow{2}{1in}{~~${K\pi\pi\pi^0}$} & ${+20~~ [14, 25]}$ &${+19~ [13, 24]}$ \\
 & ${-17 ~~[10, 23]}$ &${-17~~ [9.4, 22]}$ \\\hline
\multirow{2}{1in}{~~${\KS\pi}$} & ${+12~~ [7.3, 16]}$ &${+11~~ [7.1, 16]}$ \\
 & ${-8~~~~[4.6, 14]}$ &${-8~~~~ [4.5, 14]}$ \\\hline
\multirow{2}{1in}{~~${\KS\pi\pi^0}$} & ${+14~~ [8.9, 18]}$ &${+13~~ [8.3, 17]}$ \\
 & ${-12 ~~[6.2, 16]}$ &${-11~~ [5.6, 15]}$ \\\hline
 combined & $\pm 2.9~~ [2.1, 3.6]$ & $\pm 2.5~~ [1.5, 3.2]$ \\

\hline \multicolumn{3}{l}{$B^+ \to \Dp K^{*0}$ }\\\hline
\multirow{2}{1in}{~~${K\pi\pi}$} & ${+3.5~~ [2.5, 4.0]}$ & ${+3.3~~ [2.5, 4.0]}$ \\
 & ${-3.2 ~~[1.8, 3.6]}$ & ${-2.8~~ [1.6, 3.8]}$ \\\hline
\multirow{2}{1in}{~~${\KS\pi}$} & ${+15~~ [9.8, 19]}$ & ${+14~~ [7.9, 17]}$ \\
 & ${-11 ~~[5.8, 16]}$ & ${-7.7~ [3.8, 14]}$ \\\hline
 combined & $\pm 3.3~~ [2.1, 4.2]$ & $\pm 3.0~~ [1.8, 3.9]$ \\

\hline\hline
\end{tabular}
\end{center}
\end{table*}

\begin{table*}[htb]
\footnotesize
\renewcommand{\arraystretch}{1.2}
\begin{center}
\caption{Branching fraction fit results in units of $10^{-6}$, with
statistical uncertainties. $N_{i}$ are the yields of the fitted
species, and \BR\ represents the calculated branching fraction for
each channel.\label{tab:fit}}
\begin{tabular}{lr@{ $^+_-$ }lr@{ $\pm$ }lr@{ $\pm$ }lr@{ $^+_-$ }l}
\hline\hline
 Decay mode &\multicolumn{2}{c}{$N_{\rm sig}$} &\multicolumn{2}{c}{$N_{\BB}$} &\multicolumn{2}{c}{$N_{\rm cont}$}  &\multicolumn{2}{c}{\BR} \\
\hline \multicolumn{9}{l}{$B^+ \to \Dp K^0$}\\\hline

~~${K\pi\pi}$ & $-11.9$&$^{6.7}_{5.6}$ &$70$&$27$ &$2690$&$57$  & $-4.2$&$^{2.4}_{2.0}$ \\

~~${K\pi\pi\pi^0}$ & $10$&$^{10}_{9}$ &$111$&$51$ &$6516$&$94$  & $20$&$^{20}_{17}$ \\

~~${\KS\pi}$ & $0.6$&$^{5.3}_{4.5}$ &$20$&$14$ &$381$&$23$  & $0.7$&$^{15}_{13}$ \\

~~${\KS\pi\pi^0}$ & $-6.7$&$^{4.5}_{2.8}$ &$36$&$22$ &$1270$&$41$  & $-14$&$^{9.2}_{6.2}$\\

 combined \phantom{combi}&\multicolumn{2}{c}{-} &\multicolumn{2}{c}{-} &\multicolumn{2}{c}{-}  & $-3.4$&$^{2.2}_{1.8}$\\

\hline \multicolumn{9}{l}{$B^+ \to \Dp K^{*0}$ }\\\hline

~~${K\pi\pi}$ & $-15.6$&$^{8.7}_{7.1}$ &$463$&$63$ &$6338$&$98$  &$-5.0$&$^{2.9}_{2.1}$ \\

~~${\KS\pi}$ & $-11.4$&$ ^{3.5}_{2.4}$ &$35$&$15$ &$547$&$27$  &$-33$&$^{10.2}_{7.0}$\\

 combined &\multicolumn{2}{c}{-} &\multicolumn{2}{c}{-} &\multicolumn{2}{c}{-}  &$-5.3$&$^{2.3}_{2.0}$\\
\hline\hline
\end{tabular}
\end{center}
\end{table*}

\section{Background Characterization}
After applying the selection criteria described above, the remaining
background is composed of non-signal \BB events and continuum
events, the latter being the dominant contribution. Continuum
background events, in contrast to \BB events, are characterized by a
jet-like shape, which can be used in a Fisher discriminant
\fish~\cite{cite:Fish} to reduce this background component. The
discriminant \fish\ is a linear combination of four variables
trained to peak at 1 for signal and at $-1$ for continuum
background. The first variable is the cosine of the angle between
the $B$ thrust axis and the thrust axis of all the other
reconstructed charged tracks and neutral energy deposits (rest of
the event), where the thrust axis is defined as the direction that
maximizes the sum of the longitudinal momenta of all the particles.
The second and third variables are the event shape moments
$L_0=\sum_{i} p_i$, and $L_2 =\sum_{i} p_i |\cos \theta_i|^2$, where
the index $i$ runs over all tracks and energy deposits in the rest
of the event; $p_i$ is the momentum and $\theta_i$ is the angle with
respect to the thrust axis of the $B$ candidate. These three
variables are calculated in the CM. Finally we use $|\deltat|$, the
absolute value of the measured proper time interval between the two
$B$ decays~\cite{Aubert:2002rg}. It is calculated using the measured
separation along the beam direction $\deltaz$ between the decay
points of the reconstructed $B$ and the other $B$ and the Lorentz
boost between the laboratory and CM frames. The other $B$ decay
point is obtained from the tracks that do not belong to the
reconstructed $B$, with constraints from the reconstructed $B$
momentum and the beam-spot location. The coefficients of \fish,
chosen to maximize the separation between signal and continuum
background, are determined with samples of simulated signal and
continuum events, and
validated using off-resonance data. 
We denote two regions: the fit region, defined as $5.20<\mes<5.29
\gevcc$ and $-5< \fish <$~$5$, and the signal region, defined as
$5.27<\mes<5.29 \gevcc$ and $0< \fish <$~$5$.

To reduce the importance of the continuum background in the final
sample we divide the events according to their flavor-tagging
category~\cite{Aubert:2002rg}. We define the following exclusive
tagging categories:
\begin{itemize}
\item {\it lepton category,} events contain at least one lepton in the decay of the other \B meson;\noindent
\item{\it kaon category,} events contain at least one kaon in the decay of the other \B meson, which do not belong to the first category;\noindent
\item{\it other category} contains all the events not included in the two previous categories.\noindent
\end{itemize}
The first two categories are expected to be less contaminated by
continuum background. We fit all three categories simultaneously.
Studies of simulated events show that using the tagging categories
reduces the statistical uncertainty on the measured branching
fraction for the $K\pi\pi$ mode by 5\%, but leads to little gain for
the other modes (which are less statistically significant
themselves). Hence, we use tagging information only for the
$K\pi\pi$ channel.

The \BB background is divided into two components: non-peaking
(combinatorial) and peaking. The latter can occur when one or
several particles of a background channel are replaced by a low
momentum charged $\pi^+$ and the resulting candidate still
contributes to the signal region. The largest contributions to the
\BB\ peaking background for the $\Bp \to\ \Dp \Kz$ channel arise
from the following decays: $\Bzb\to \Dp\rho^-$ with $\Dp$ decaying
into signal channels, $B^0\to \Dzb K^0$ and $B^0\to \Dstarzb K^{0}$.
To further reduce the contribution from the $\Bzb\rightarrow
\Dp\rho^-$ background, the variable $|\cos\theta_{\KS}|$ has been
introduced, where $\theta_{\KS}$ is the \KS helicity angle, i.e.,
the angle between one of the two pions from the \KS and the \Dp in
the \KS rest frame. We reject events with $|\cos\theta_{\KS}|$
greater than 0.8 for the $K \pi \pi$ mode and 0.9 for all other
modes. Based on MC studies, we expect no more than 1 \BB peaking
background event per mode in the signal region, after applying all
selection criteria (see Table~\ref{tab:eff}). A similar study is
performed for the $\Bp \to\ \Dp \Kstarz$ decay modes. The main
peaking backgrounds arise from $\Bzb\to D^+\rho^-$, $\Bzb\to
D^+K^{*-}$, and $\Bzb\to D^+a_1^{-}$. In all cases, the $D^+$ decays
into the signal decay modes. The number of \BB peaking background
events expected in the signal region for the $DK^*$ mode are shown
in Table~\ref{tab:eff}.

Charmless $B$ decays may also contribute to the peaking background.
These decays can produce $\pi$ and $K$ mesons with characteristics
similar to those of signal events without forming a real $D$ meson.
The charmless background is evaluated from data using the \Dp
sidebands: events are required to satisfy the criteria
$1.774<M_D<1.840 \gevcc$ or $1.900<M_D<1.954 \gevcc$. We obtain $-1.7
\pm 1.0$ events for $D K$ decays and $-0.7 \pm 2.1$ events for
$DK^*$ decays. We estimate the charmless peaking background
contribution to be negligible and assign a systematic uncertainty based on
this assumption.

The overall reconstruction and selection efficiencies for signal
events, as well as the numbers of expected events for each
background category, are given in Table~\ref{tab:eff}.

\section{Fit Procedure}

The signal and background yields are extracted by maximizing the
unbinned extended likelihood
\begin{equation}
\mathcal{L} = (e^{- N'}/N !)\cdot {N'}^{N} \cdot \prod_{j=1}^{N}
f({\bf x}_{j} \mid {\bf \theta}, N').
\end{equation}
Here ${\bf x}_j = \{\mes;\fish\}$, $\bf \theta$ is a set of
parameters, $N$ is the number of events in the selected sample, $N'$
is the expectation value for the total number of events, and
\begin{equation}
f({\bf x} \mid {\bf \theta}, N')=\frac{N_{\rm sig}f_{\rm sig}({\bf
x}| {\bf \theta}) + \sum_{i}N_{B_{i}} f_{B_{i}}({\bf x}|{\bf
\theta})}{N'},
\end{equation}
with $f_{\rm sig}({\bf x}| {\bf \theta})$ and $f_{B_{i}}({\bf
x}|{\bf \theta})$ the probability density functions (PDFs) for the
hypothesis that the event is a signal or a background event,
respectively. The $B_{i}$ are the different background categories
used in the fit: continuum background, combinatorial \BB background,
and peaking \BB background. $N_{\rm sig}$ is the number of signal
events, and $N_{B_i}$ is the number of events for each background species
$B_{i}$.

The individual probability density functions are defined by the
product of the one-dimensional distributions of \mes\ and \fish.
Absence of the correlations between these distributions is checked
using the MC samples. The signal \mes distribution is modeled with a
Gaussian function. The continuum and non-peaking \BB background
\mes\ distributions are modeled with two different threshold ARGUS
functions defined~\cite{cite:argus} as follows:
\begin{equation}
A(x) = x\sqrt{1-\left(\frac{x}{x_0}\right)^2}\cdot
e^{c\left(1-\left(\frac{x}{x_0}\right)^2\right)},\label{eq:threshold}
\end{equation}
where $x_0$ represents the maximum allowed value for the variable
$x$ and $c$ accounts for the shape of the distribution. The \mes
distribution of the peaking \BB background is modeled with a Crystal
Ball (CB) function \cite{cite:CB}. The CB function is a Gaussian
modified to include a power-law tail on the low side of the peak.
The \fish\ distributions are modeled as the sum of two asymmetric
Gaussians for signal and continuum background events, and with a
Gaussian for the combinatorial \BB background. For the peaking \BB
background we use a Gaussian distribution for the $DK$ mode. For the
$DK^*$ mode, an asymmetric Gaussian is used for the $K\pi\pi$ mode
and a sum of two asymmetric Gaussians for the $\KS\pi$ mode. The
shape parameters of the threshold function for continuum background
are determined from data. All other PDF parameters are derived from
the simulated events.

In the fits we fix the numbers of peaking \BB\ background events,
which are estimated from the PDG branching fractions~\cite{cite:PDG}
and MC efficiency evaluations.

The number of signal events determined by the fit ($N_{\rm sig}$) is
used to calculate the branching fraction as
\begin{equation}
\label{eq:BrR} \BR(B^{+} \to \Dp K^{0}) =\notag\frac{N_{\rm
sig}}{N_{\Bp}\cdot \epsilon_{\rm sig}}\cdot\frac{2}{\BR_{D}\cdot
\BR_{\KS}},
\end{equation}
\noindent where $N_{\Bp}$ is the total number of charged $B$ mesons
in the data sample (equal to the total number of all \BB pairs
produced, since we assume equal production of \BpBm and \BzBzb),
$\BR_{D}$ and $\BR_{\KS}$ are the branching fraction for each $D$
meson decay channel and for $\KS\to\pip\pim$
respectively~\cite{cite:PDG}, and $\epsilon_{\rm sig}$ is the
reconstruction efficiency for each $D$ decay channel evaluated from
MC events. The expression for $\BR(B^{+} \to \Dp K^{*0})$ is
obtained replacing $\BR_{\KS}/2$ with the branching fraction of
$\Kstarz\to\Kp\pim$, $\BR_{\Kstarz}$. The likelihoods for individual
channels are combined to derive average branching fractions for
$\Bp\to\Dp\Kz$ and $\Bp\to\Dp\Kstarz$.

The fit procedure is validated using an ensemble of simulated
experiments (toy MC studies) with all yields generated according to
Poisson distributions. The non-floating parameters of the fits as
well as the shapes of the background threshold functions are fixed
to the values obtained from the MC samples. We define the pull for a
variable $x$ as the difference between the fitted $x_{\rm fit}$ and
the mean generated value $\langle x_{\rm gen}\rangle$, divided by
the error $\sigma_{\rm err}$, $x_{\rm pull}=(x_{\rm fit}-\langle
x_{\rm gen}\rangle)/\sigma_{\rm err}$. We use the negative errors
for fitted values that are smaller than the generated ones and the
positive errors in the opposite case. The procedure gives
Gaussian-like pull distributions for each channel and thus no biases
of the fit model were found.  In Table \ref{tab:errors} we show
resulting expectations of asymmetric errors for each channel. The
95\% probability ranges for these errors obtained from toy MC
studies are also shown. Tests of the fit procedure performed on the
full MC samples give values for the yields compatible with the
generated ones.

\begin{figure*}[h!]
\begin{center}
\setlength{\unitlength}{1\linewidth}
\begin{picture}(0.3,0.15)
\epsfig{file=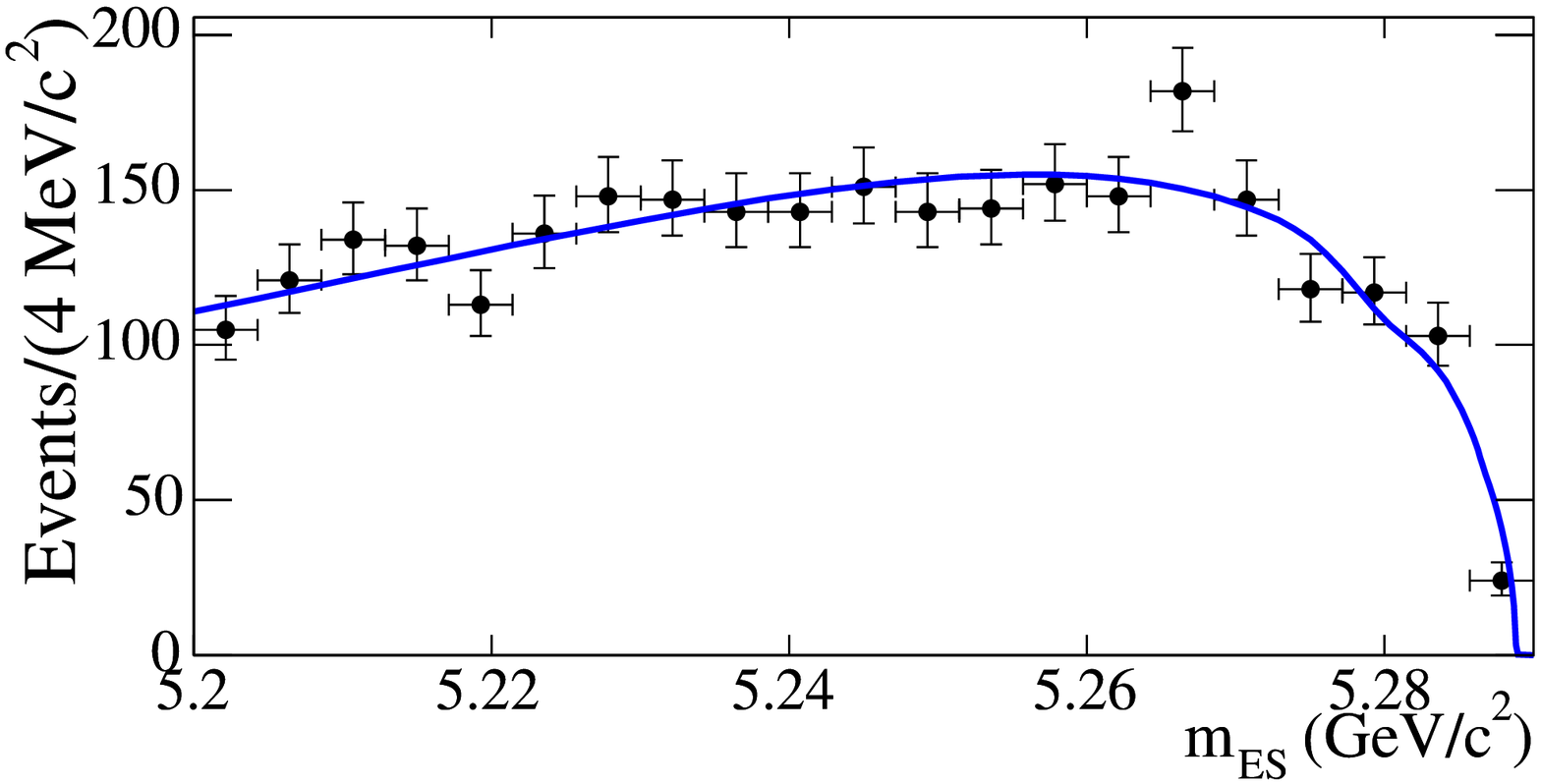,width=0.3\linewidth}
\put(-0.25,0.035){(a)}
\end{picture}
\begin{picture}(0.3,0.15)
\epsfig{file=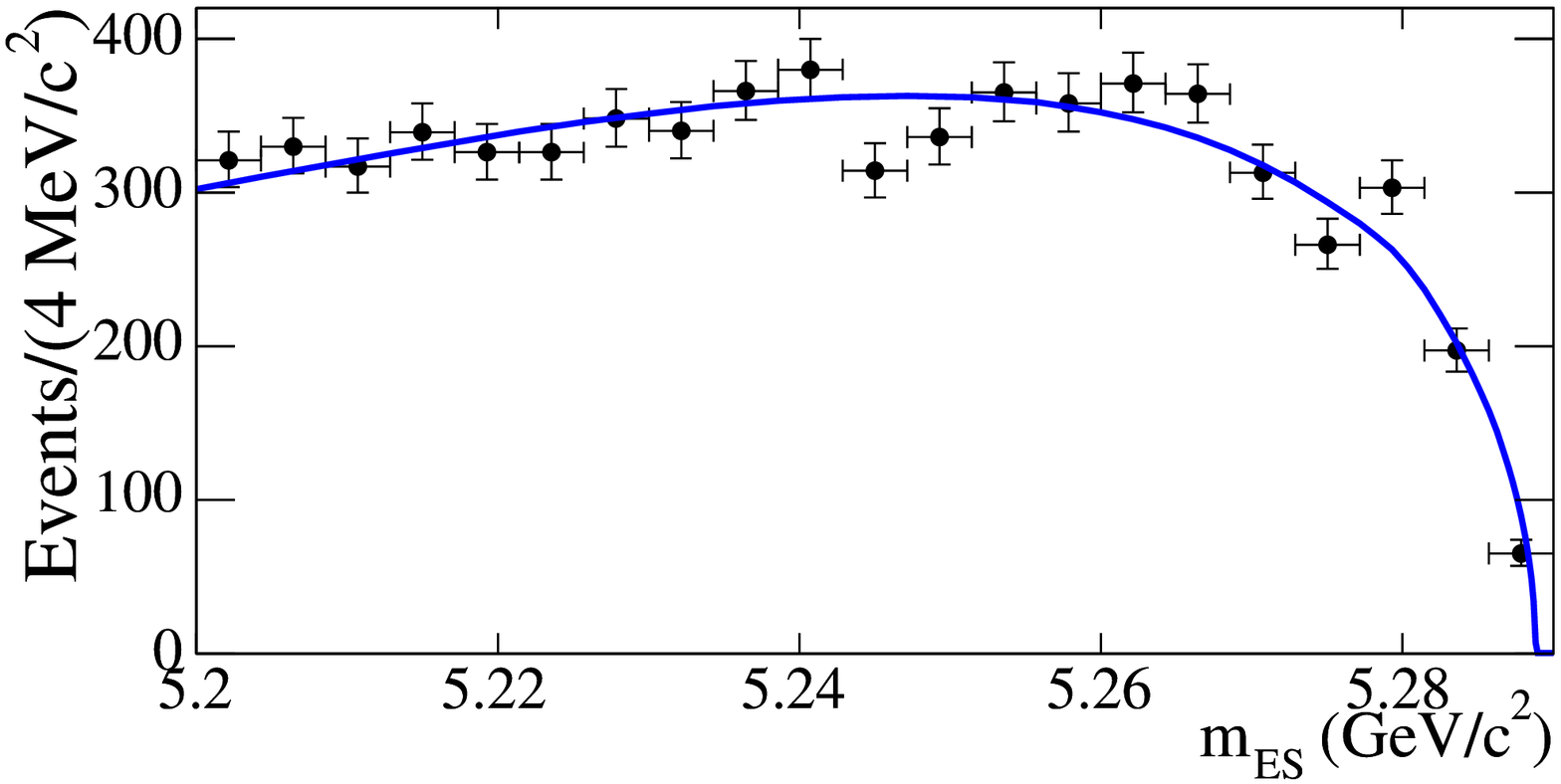,width=0.3\linewidth}
\put(-0.25,0.035){(b)}
\end{picture}
\begin{picture}(0.3,0.15)
\epsfig{file=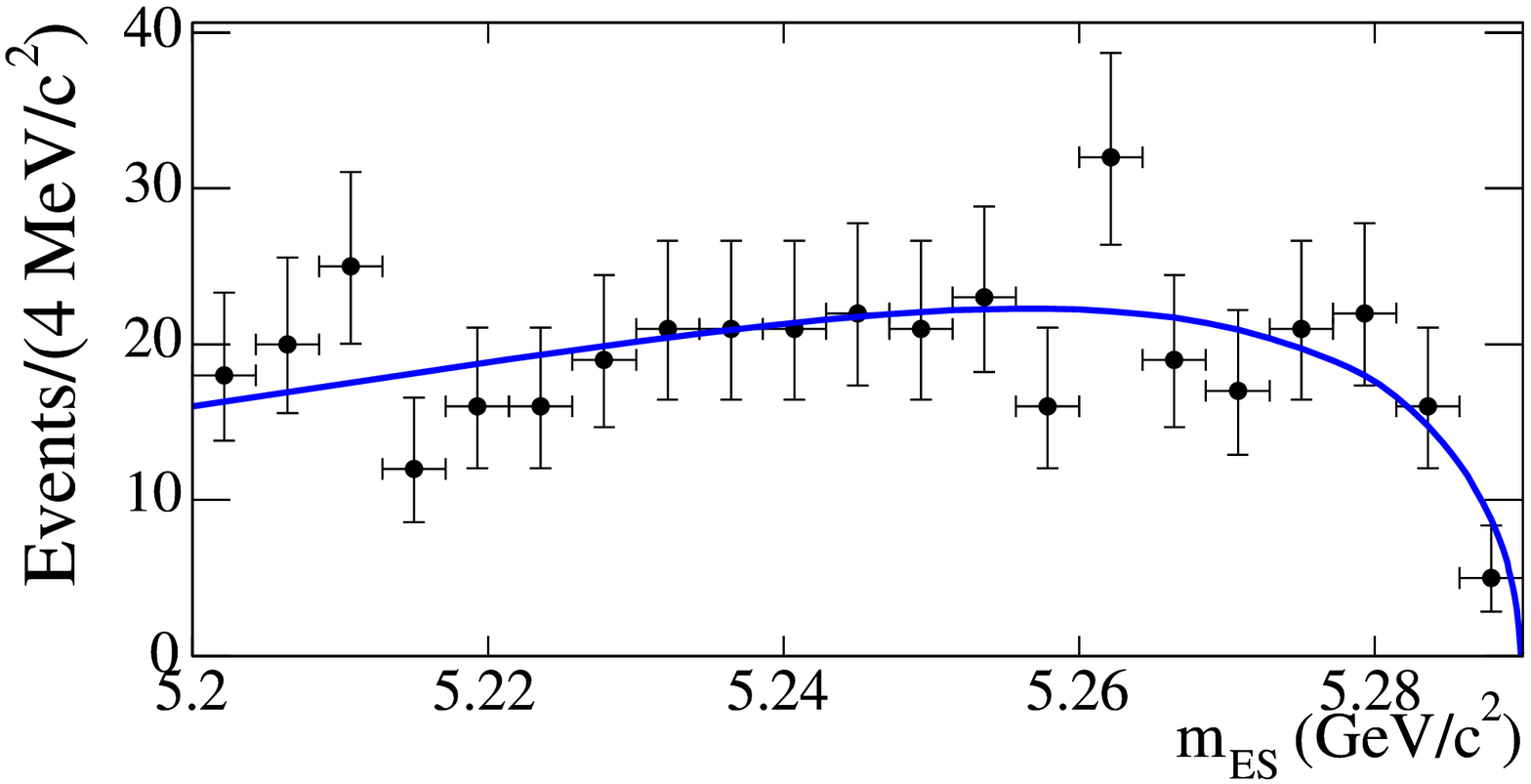,width=0.3\linewidth}
\put(-0.25,0.035){(c)}
\end{picture}\\
\begin{picture}(0.3,0.15)
\epsfig{file=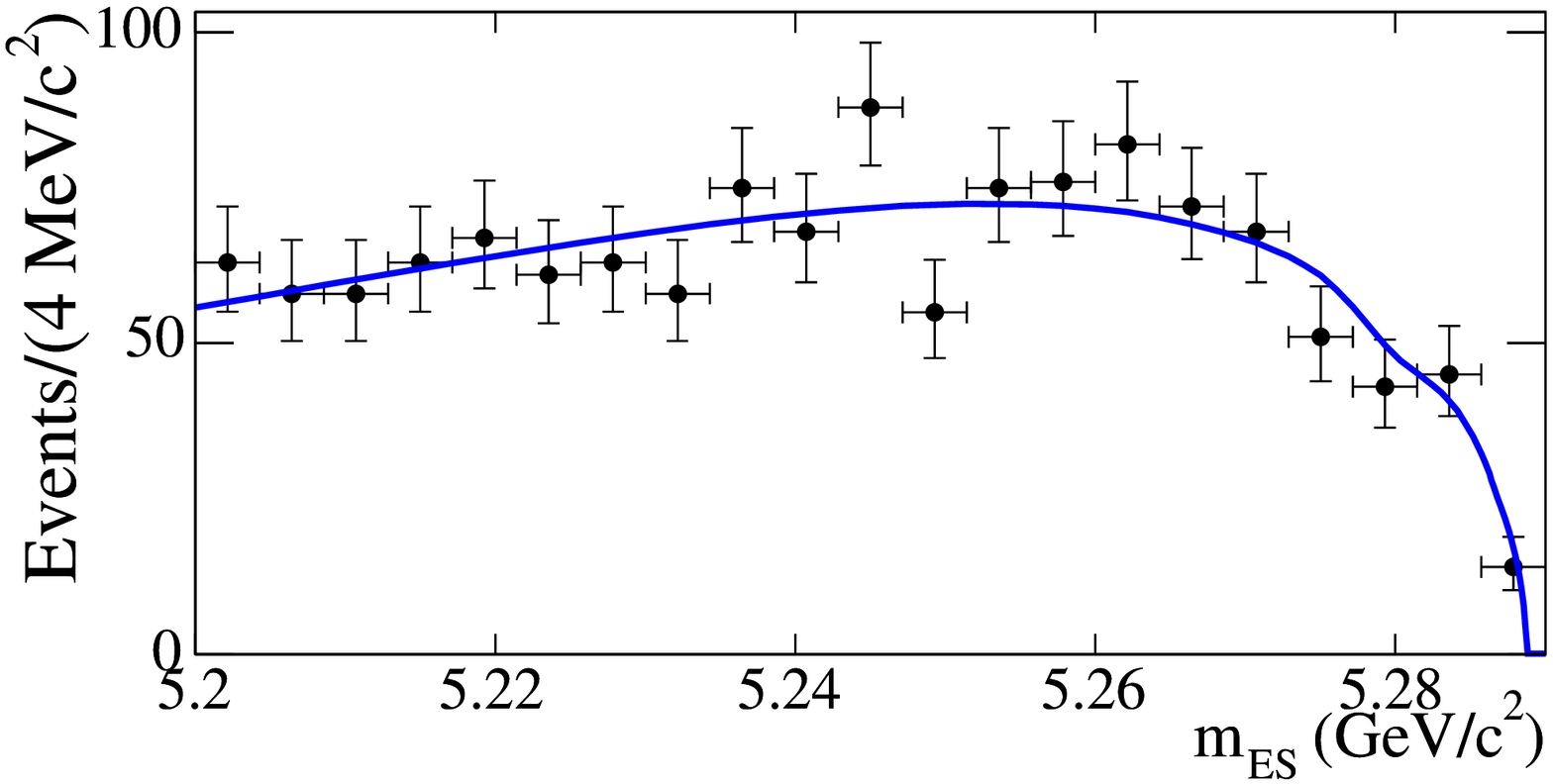,width=0.3\linewidth}
\put(-0.25,0.035){(d)}
\end{picture}
\begin{picture}(0.3,0.15)
\epsfig{file=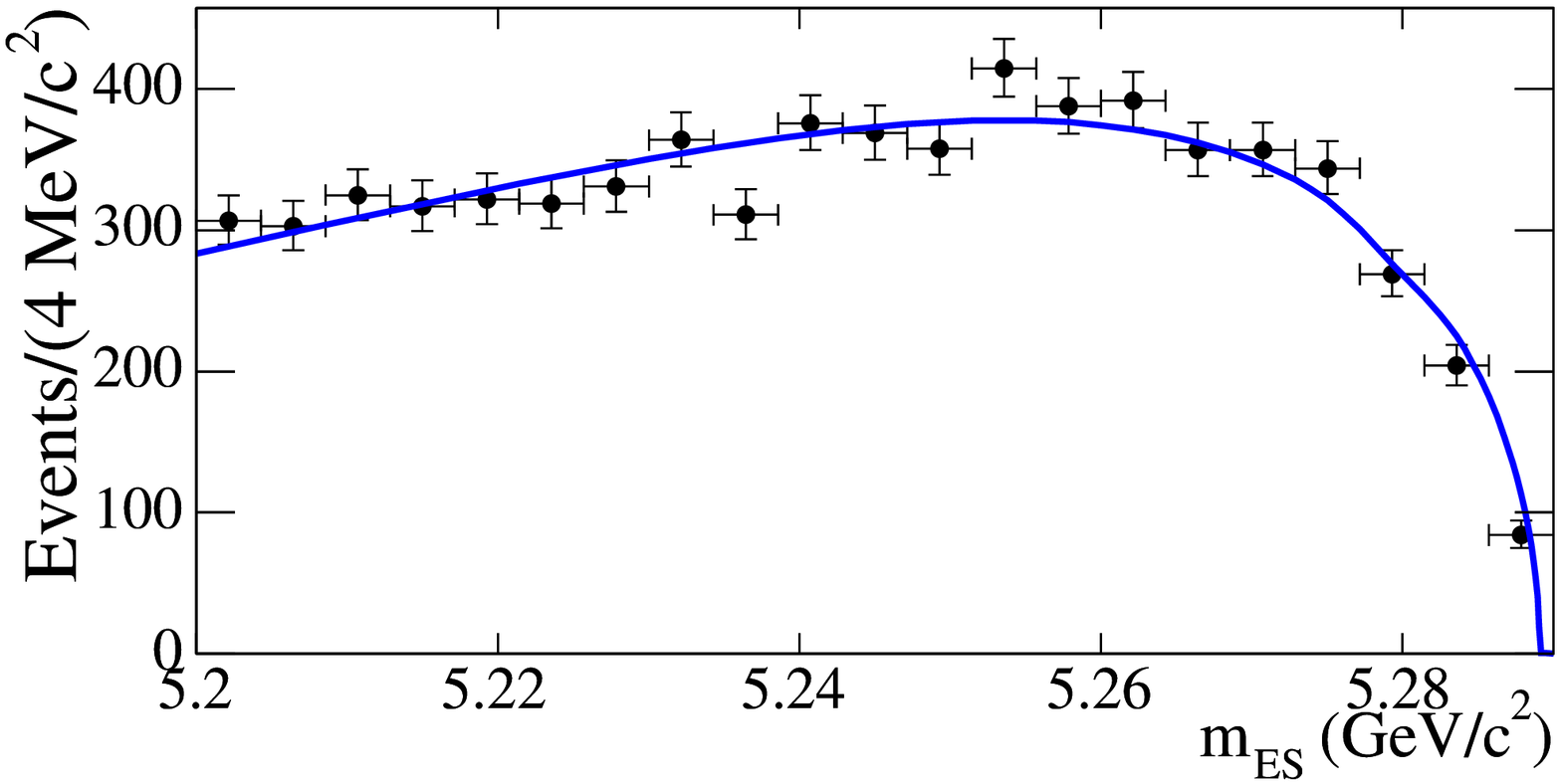,width=0.3\linewidth}
\put(-0.25,0.035){(e)}
\end{picture}
\begin{picture}(0.3,0.15)
\epsfig{file=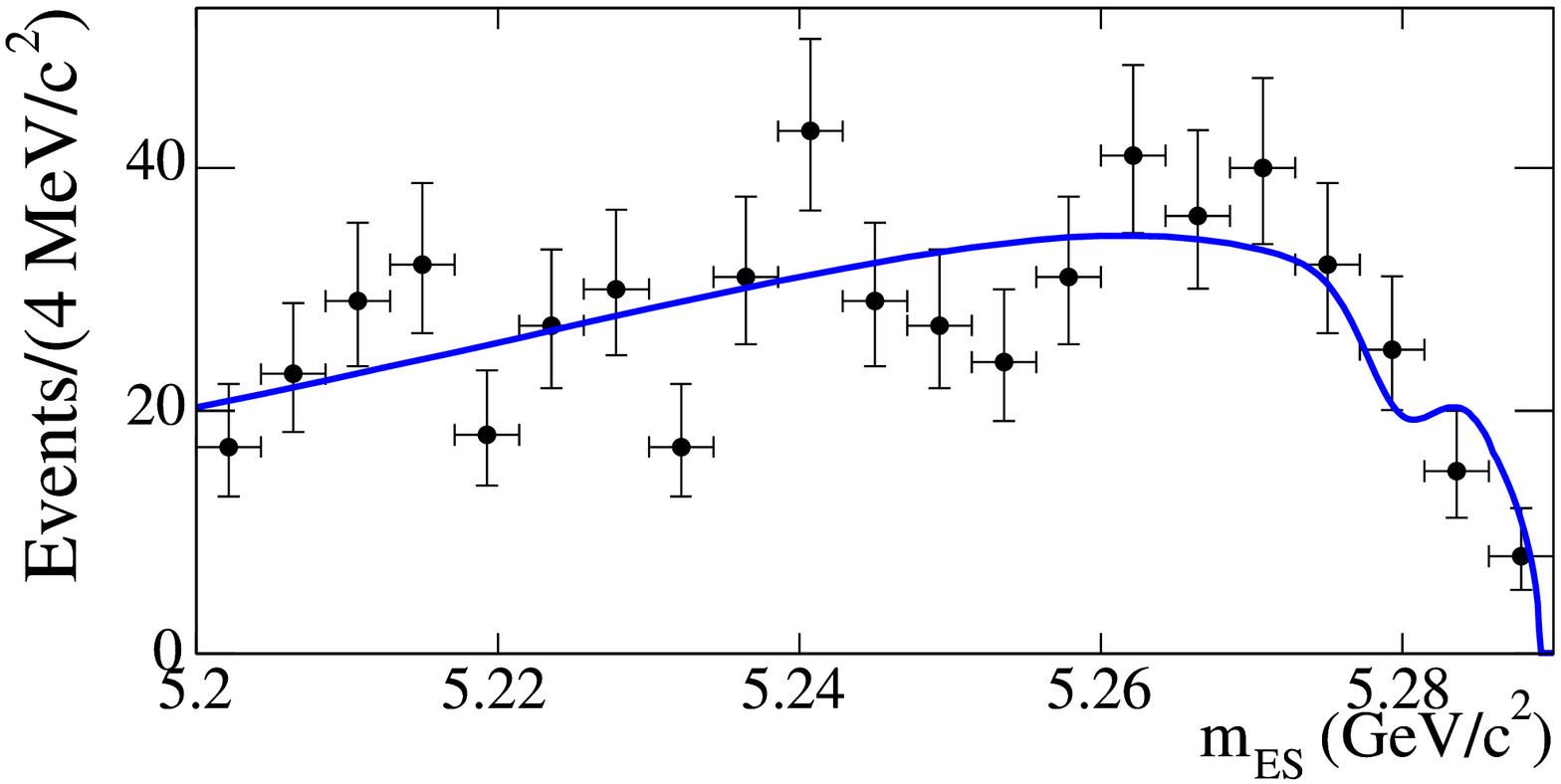,width=0.3\linewidth}
\put(-0.25,0.035){(f)}
\end{picture}
\begin{picture}(0.3,0.15)
\epsfig{file=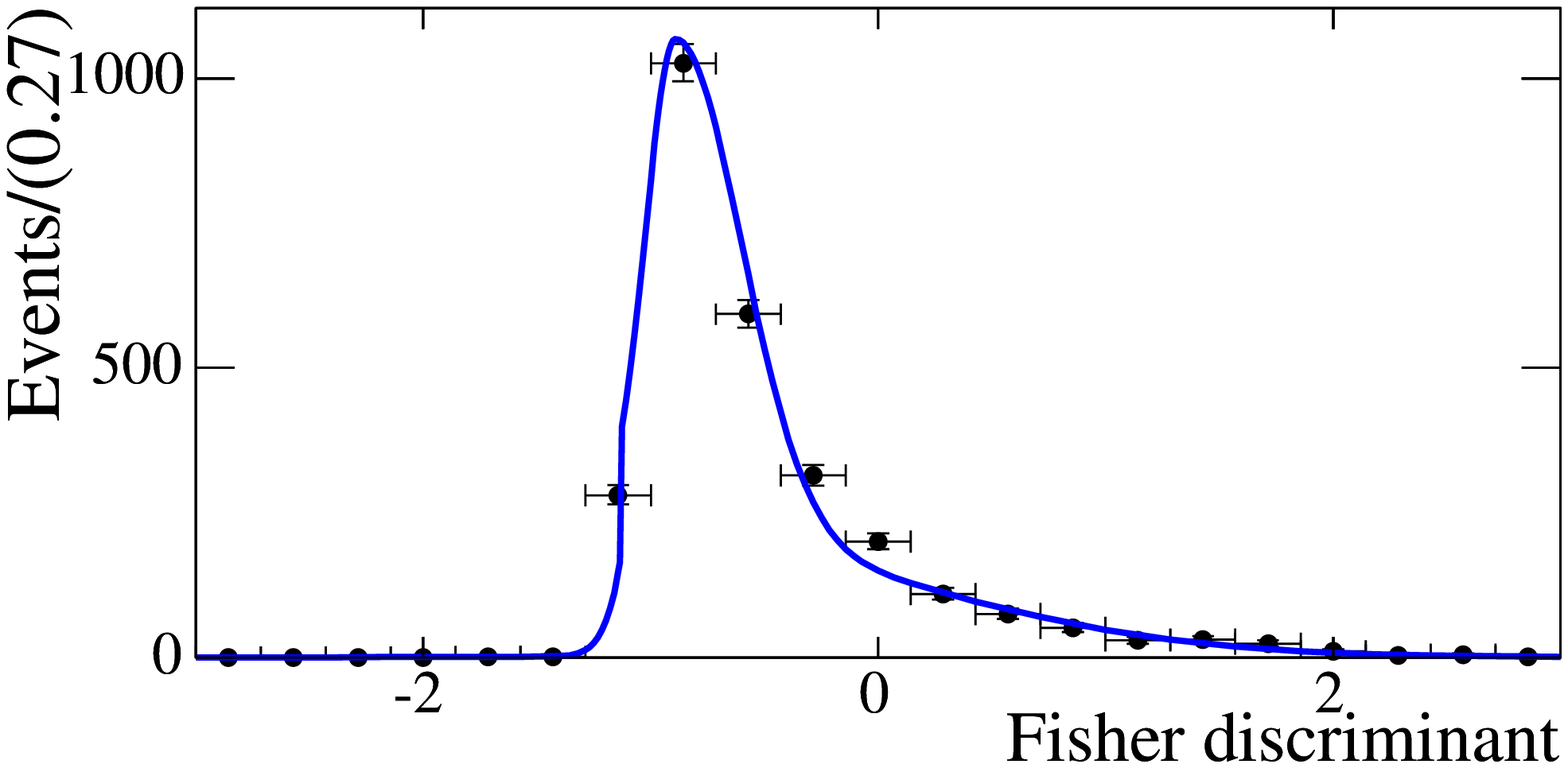,width=0.3\linewidth}
\put(-0.25,0.115){(a)}
\end{picture}
\begin{picture}(0.3,0.15)
\epsfig{file=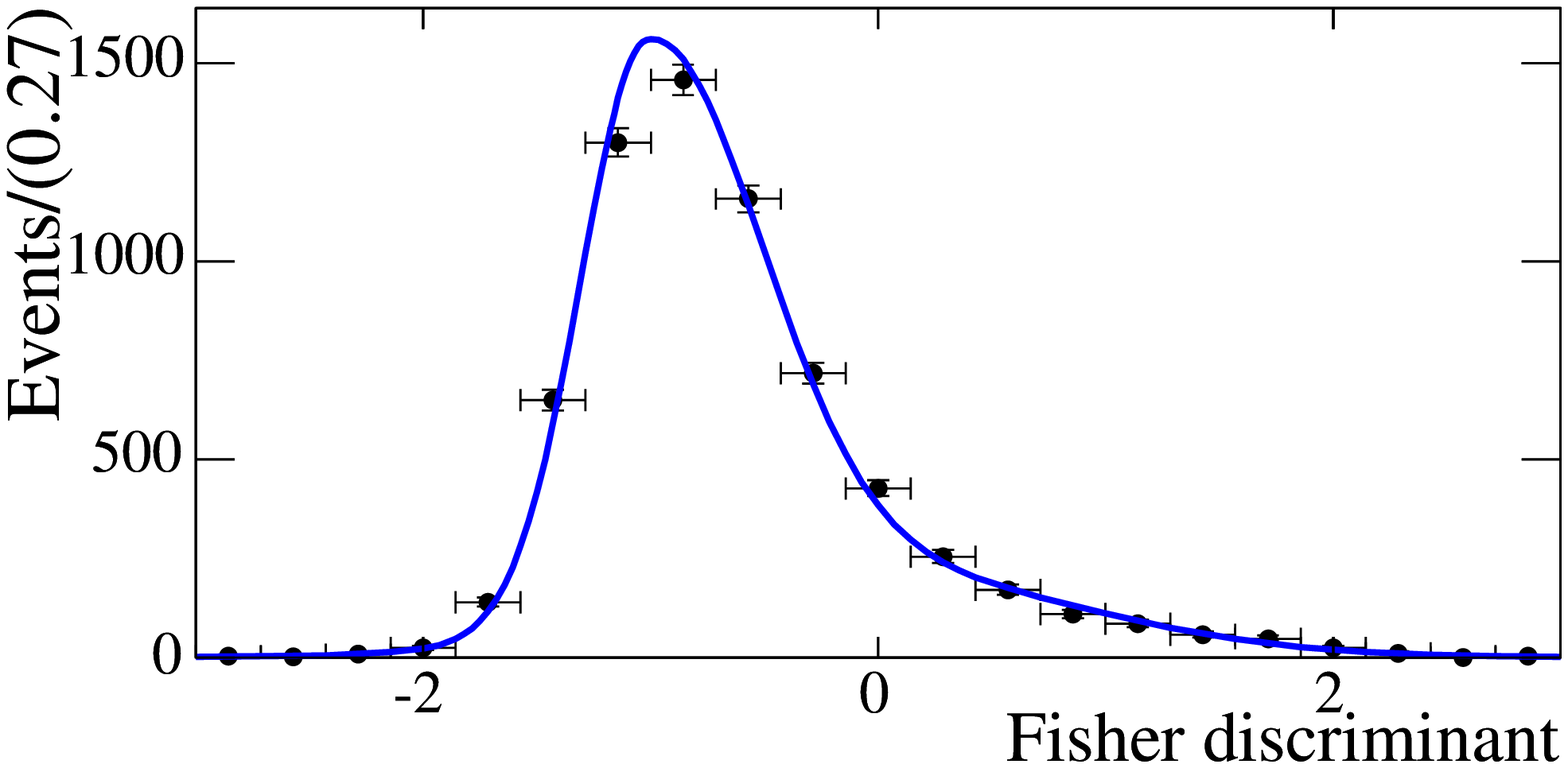,width=0.3\linewidth}
\put(-0.25,0.115){(b)}
\end{picture}
\begin{picture}(0.3,0.15)
\epsfig{file=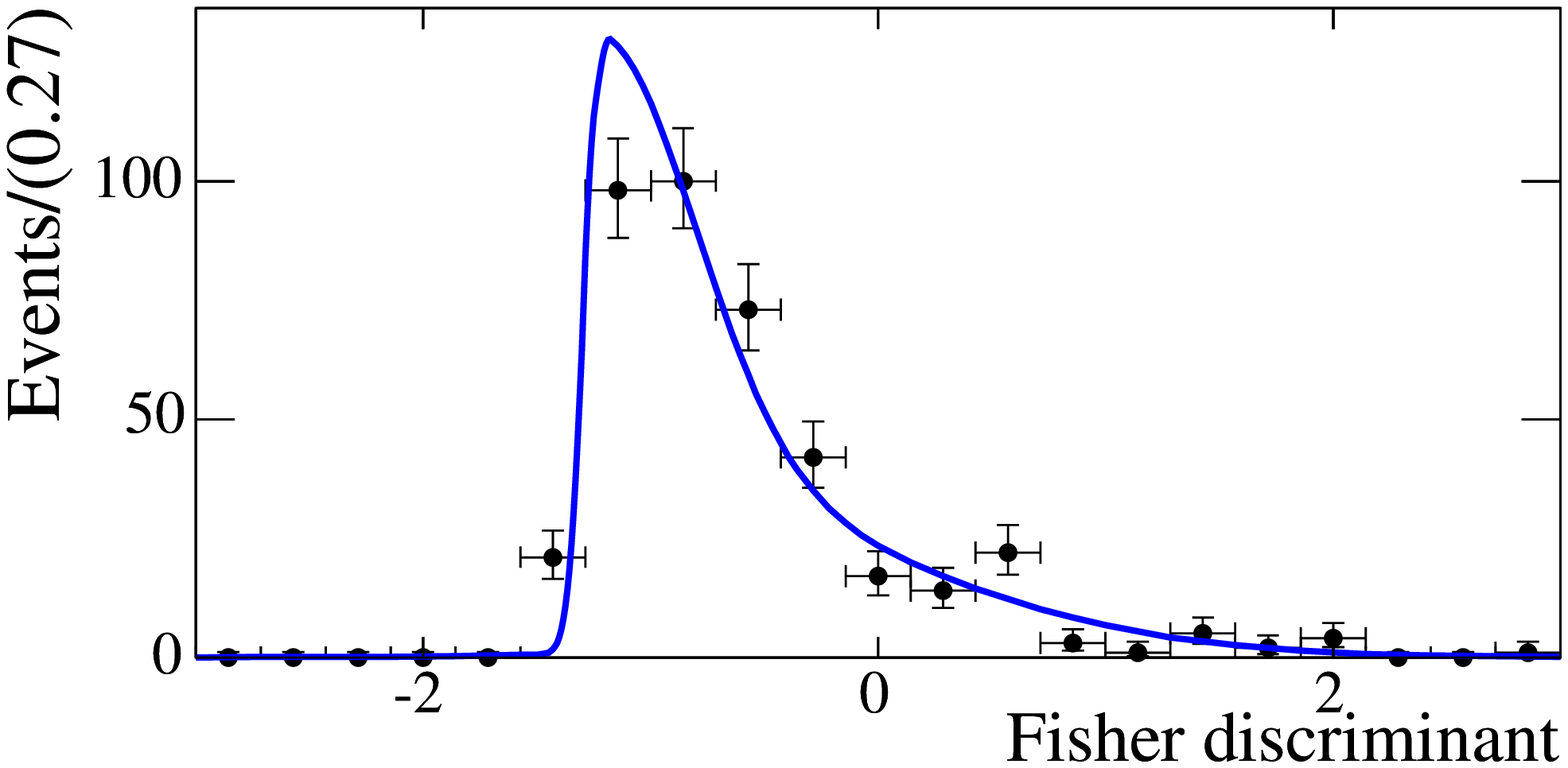,width=0.3\linewidth}
\put(-0.25,0.115){(c)}
\end{picture}\\
\begin{picture}(0.3,0.15)
\epsfig{file=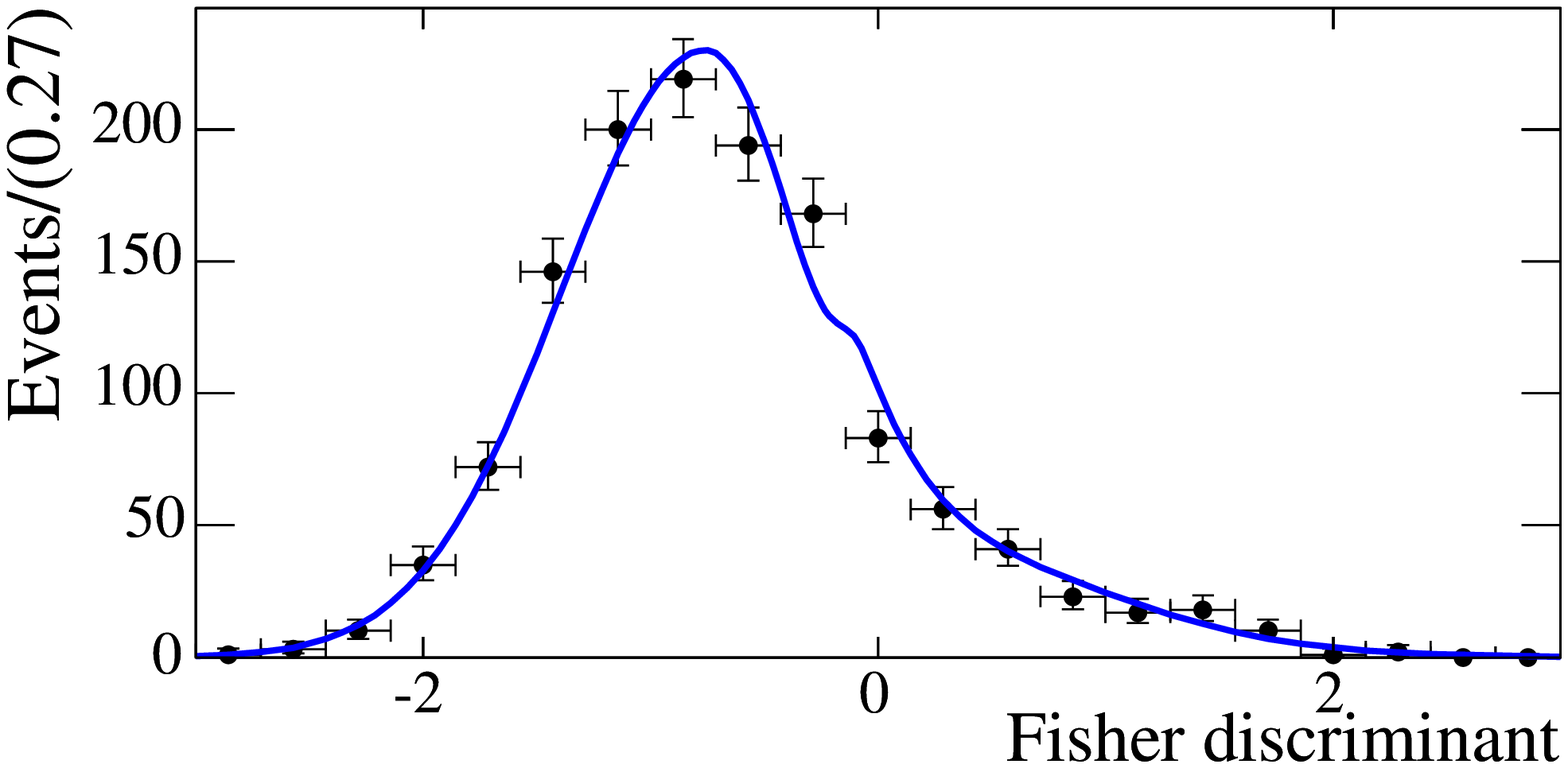,width=0.3\linewidth}
\put(-0.25,0.115){(d)}
\end{picture}
\begin{picture}(0.3,0.15)
\epsfig{file=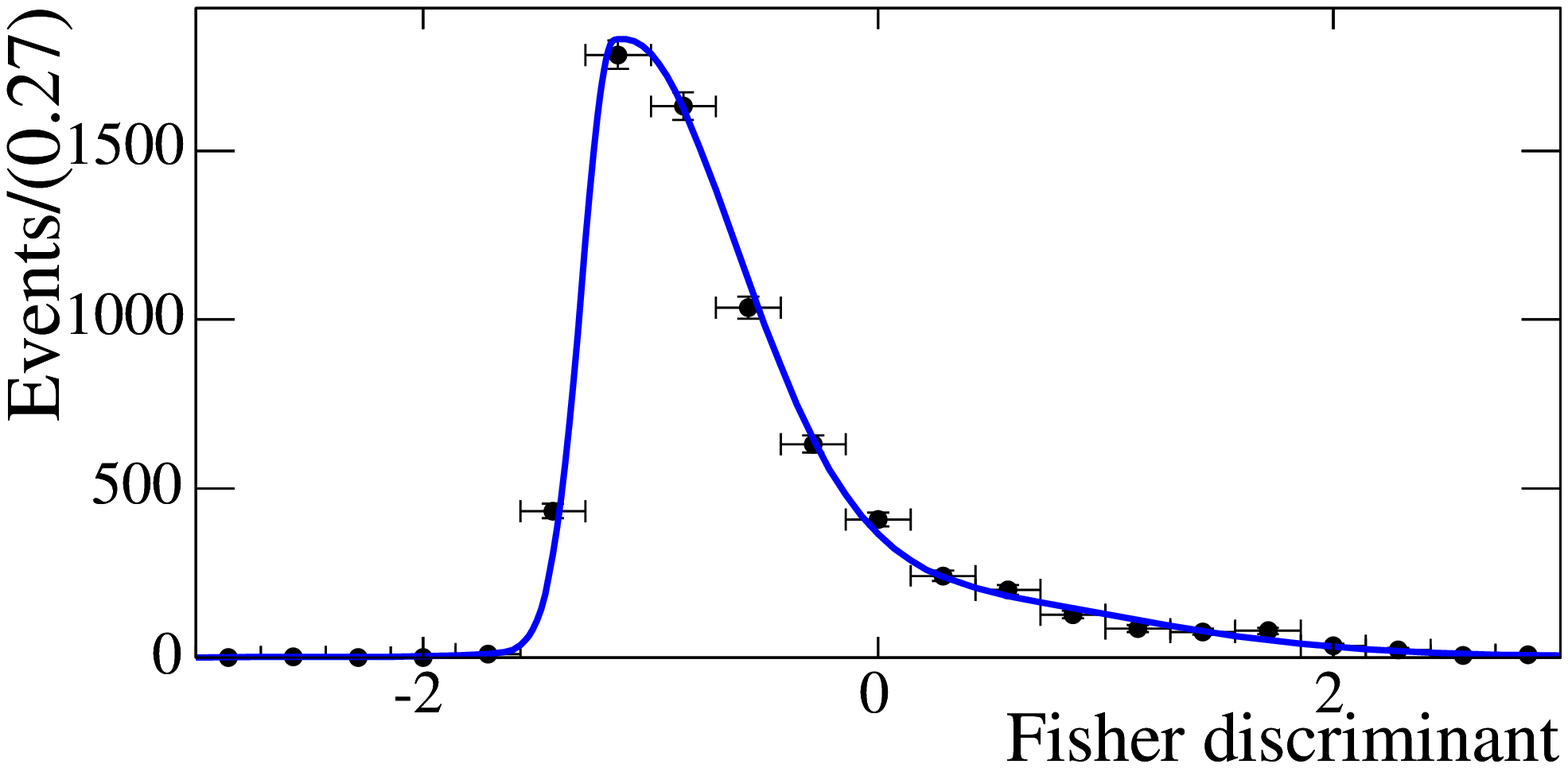,width=0.3\linewidth}
\put(-0.25,0.115){(e)}
\end{picture}
\begin{picture}(0.3,0.15)
\epsfig{file=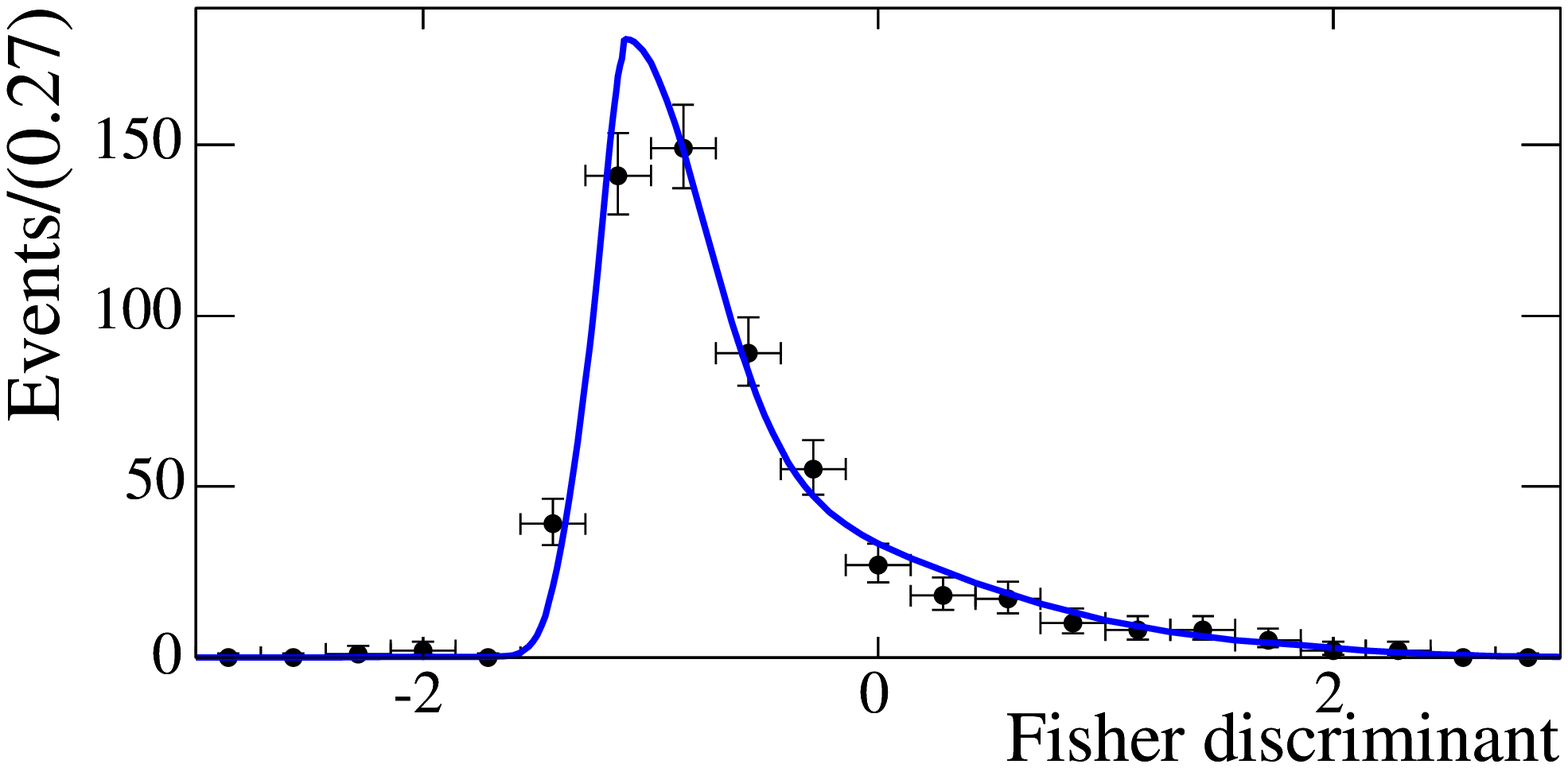,width=0.3\linewidth}
\put(-0.25,0.115){(f)}
\end{picture}
\caption{(color online) Projections of the 2D likelihood function
onto the \mes\ (top two rows) and \fish\ (bottom two rows) axes for
(a)~$K \pi \pi$, (b)~$K \pi \pi \pi^0$, (c)~$\KS \pi$ and (d)~$\KS
\pi \pi^0$ for the $\Bp \to\ \Dp \KS$ mode, and (e)~$K \pi \pi$ and
(f)~$\KS \pi$ for the $\Bp \to\ \Dp \Kstarz$ mode. The data are
indicated with black dots and error bars and the (blue) solid curve
is the projection of the fit.\label{fig:proj_fisher}}
\end{center}
\end{figure*}

\begin{figure*}[htb]
\begin{center}
\setlength{\unitlength}{1\linewidth}
\begin{picture}(0.3,0.15)
\epsfig{figure=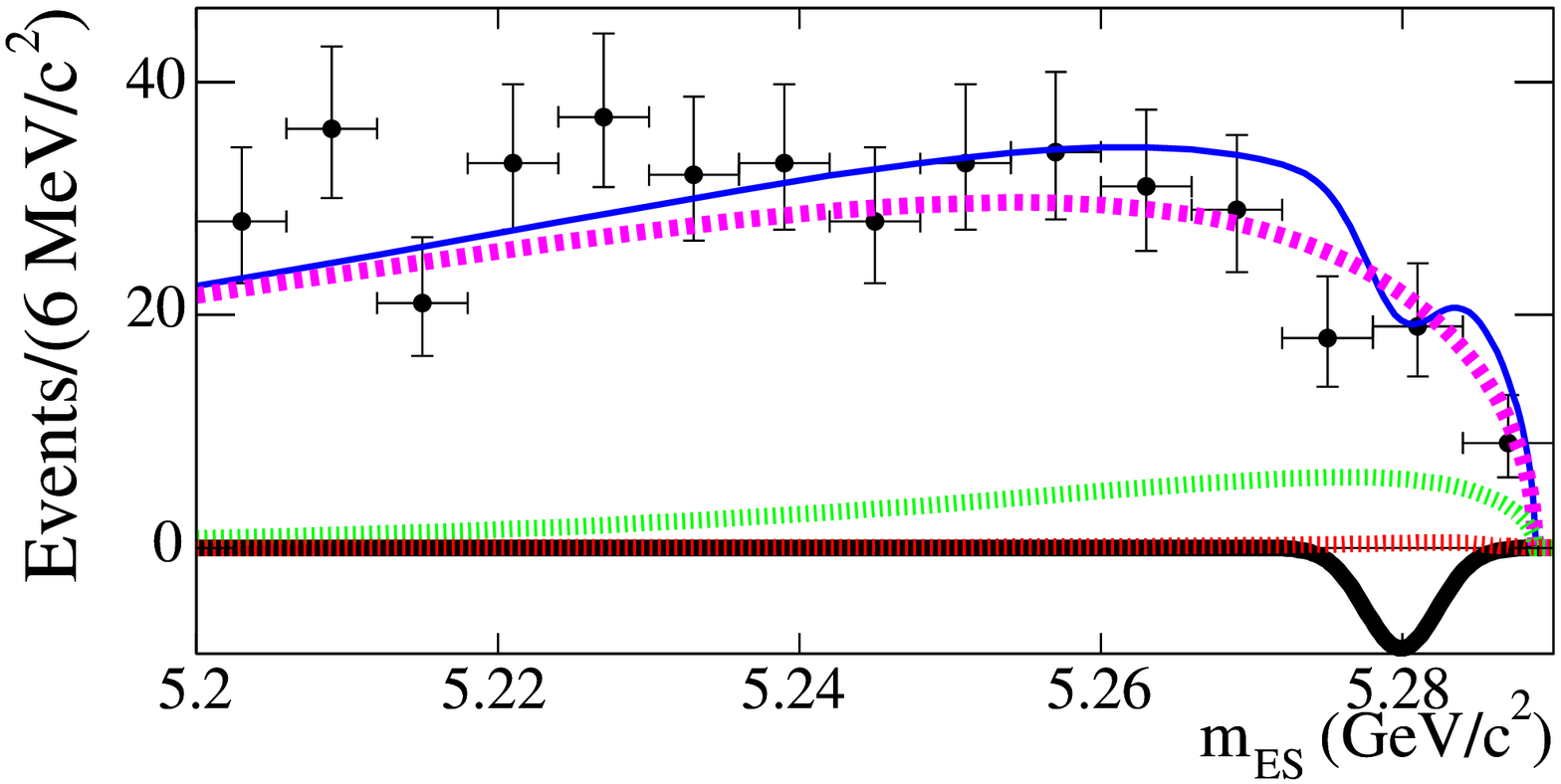,width=0.3\linewidth}
\put(-0.25,0.065){(a)}
\end{picture}
\begin{picture}(0.3,0.15)
\epsfig{figure=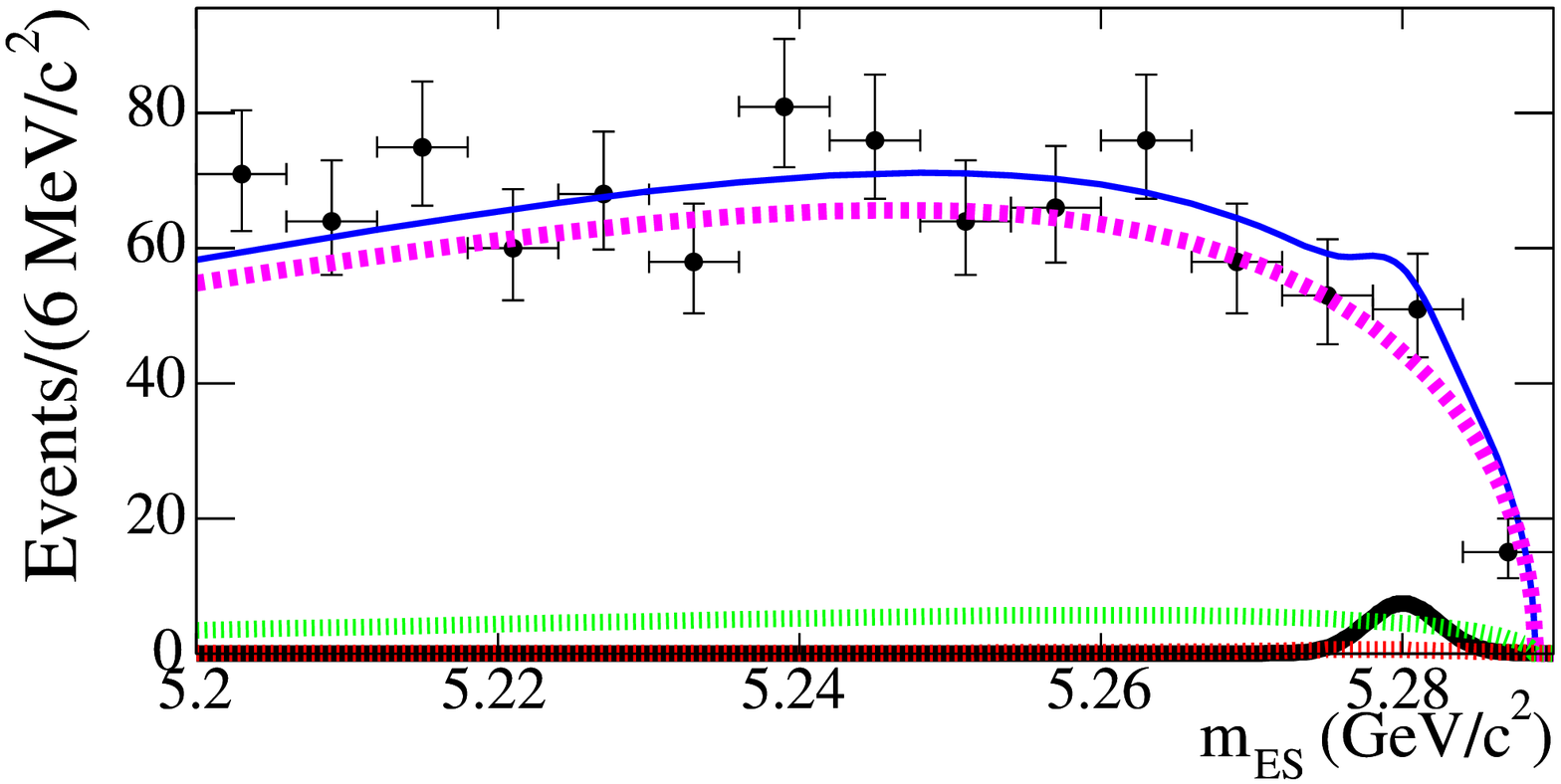,width=0.3\linewidth}
\put(-0.25,0.065){(b)}
\end{picture}
\begin{picture}(0.3,0.15)
\epsfig{figure=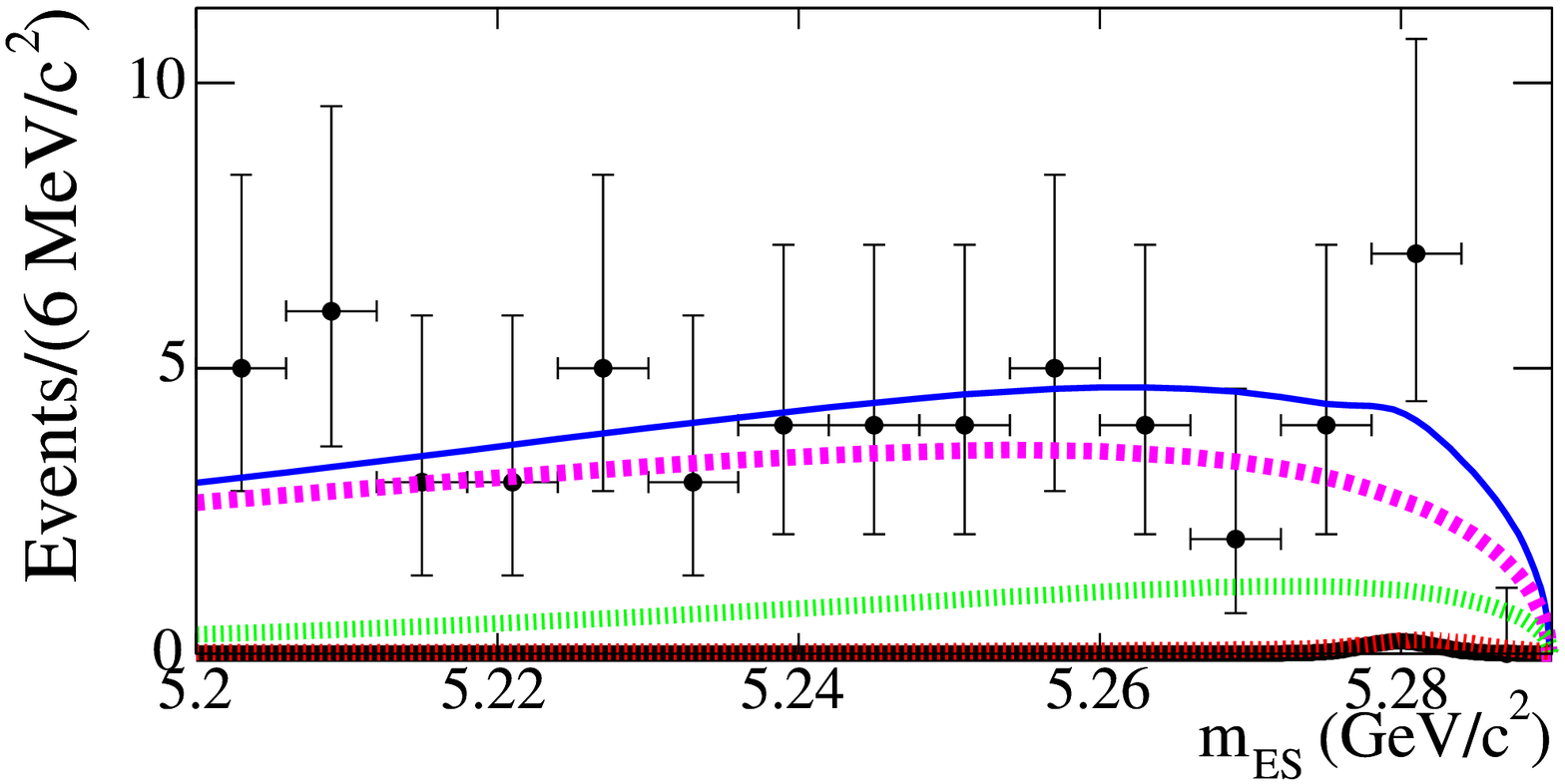,width=0.3\linewidth}
\put(-0.17,0.115){(c)}
\end{picture}\\
\begin{picture}(0.3,0.15)
\epsfig{figure=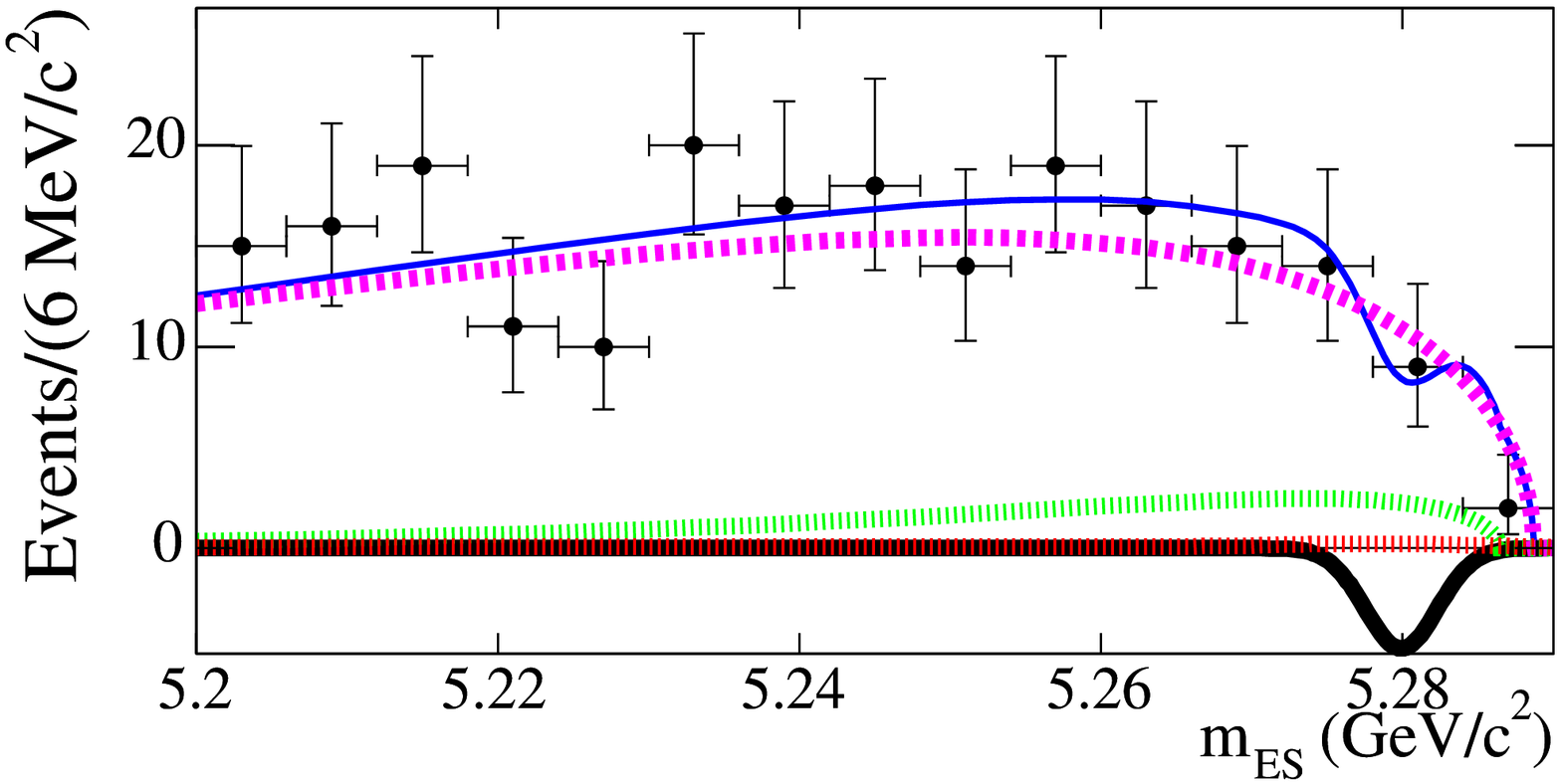,width=0.3\linewidth}
\put(-0.25,0.065){(d)}
\end{picture}
\begin{picture}(0.3,0.15)
\epsfig{figure=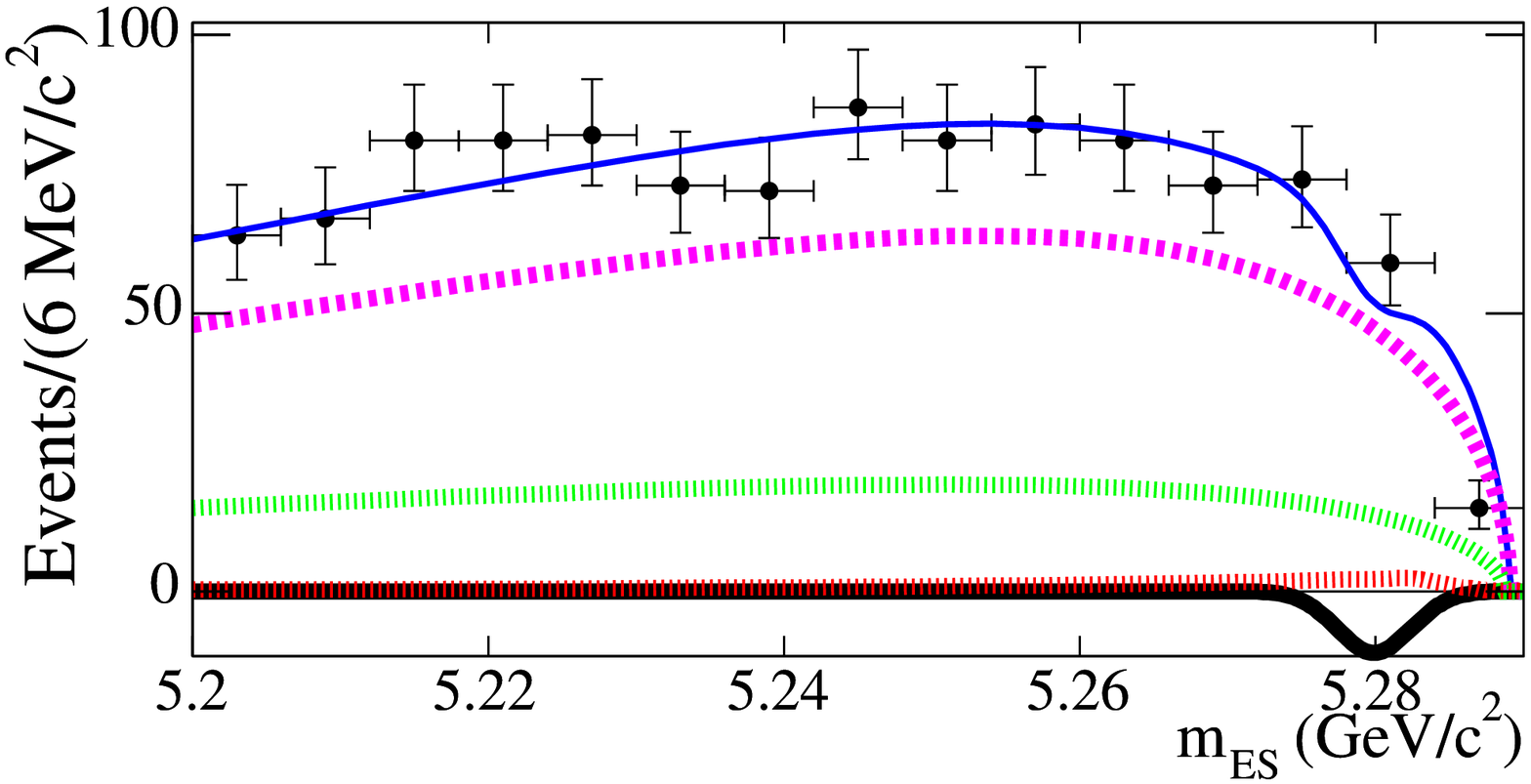,width=0.3\linewidth}
\put(-0.25,0.065){(e)}
\end{picture}
\begin{picture}(0.3,0.15)
\epsfig{figure=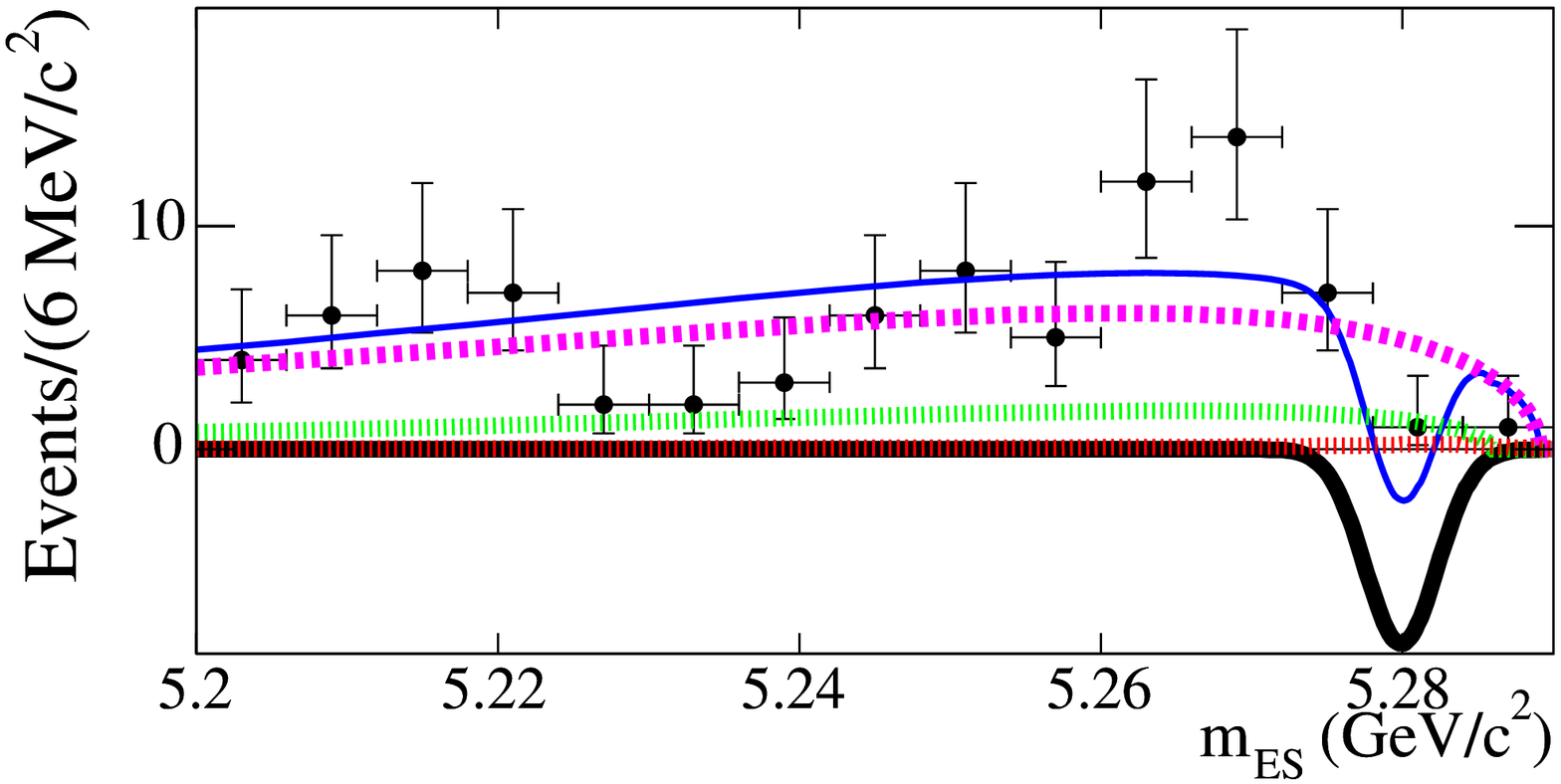,width=0.3\linewidth}
\put(-0.25,0.035){(f)}
\end{picture}
\caption{(color online) From top left to bottom right: \mes\
projection for (a)~$K \pi \pi$, (b)~$K \pi \pi \pi^0$, (c)~$\KS
\pi$, and (d)~$\KS \pi \pi^0$ for the $\Bp \to\ \Dp \KS$ mode and
(e)~$K \pi \pi$ and (f)~$\KS \pi$ for the $\Bp \to\ \Dp \Kstarz$
mode. The data are indicated with black dots and error bars and the
different fit components are shown: signal (black solid curve),
combinatorial \BB (green dotted), continuum (magenta dot-dashed) and
\BB\ peaking background (red dotted) and the blue solid curve is the
projection of the fit. We require $\fish$~$>0$ to visually enhance
the signal component. Such a cut has an approximate efficiency of
70\% for signal, while it rejects more than 80\% of the continuum
background.\label{fig:shapes_dkst}}
\end{center}
\end{figure*}

The main results of the fit to the data are reported in
Table~\ref{tab:fit}, which gives the values of the fitted parameters
for each $D$ channel and for the combination of fits. The background
yields are close to the expectations and the errors obtained on the
branching fractions are in good agreement with the values reported
in Table~\ref{tab:errors}. The leading contribution (as expected) is
obtained from the $K\pi\pi$ mode. Likelihood fit projections of the
\mes\ and \fish\ distributions are shown in
Fig.~\ref{fig:proj_fisher}. In Fig.~\ref{fig:shapes_dkst} we also
show for illustrative purposes the fit projection for \mes, after
requiring $\fish$~$>0$, to visually enhance any possible signal.

\section{Systematic errors}

We consider various sources of systematic error. One of the largest
contributions comes from the uncertainties on the PDF
parameterizations. To evaluate the contributions related to the
$m_{\rm ES}$ and \fish\ PDFs, we repeat the fit varying the
MC-obtained PDF parameters within their statistical errors, taking
into account correlations among the parameters (labeled as ``PDF -
MC'' in the final list of systematic error sources). Differences
between the data and MC (labeled as ``Data - MC PDF shapes'' in the
final list of systematic error sources) for the shapes of \mes\ and
\fish\ distributions are studied for signal components using data
control samples. $\Bzb \to D^+ \pi^-$ and $\Bzb \to \Dp \rho^-$
selected events are used to obtain the \mes\ and \fish\ parameters
for the $DK$ and $DK^*$ modes, respectively. The analysis strategy
is the same as for the signal events except for specific criteria to
select \KS or \Kstarz. For the continuum background, we estimate
this uncertainty by repeating the fit using the PDF parameters
obtained from off-resonance data instead of those from continuum MC.
Finally, for the \BB\ background, we estimate this uncertainty by
leaving the parameters that describe the \BB combinatorial
background as free variables in the fit (separately for \mes\ and
\fish). The systematic uncertainty is defined as the difference in
the branching fraction results from the nominal and alternative fits
summed in quadrature.

The systematic errors on the signal reconstruction efficiency
include the uncertainty due to limited MC statistics, uncertainties
on possible differences between data and MC in tracking efficiency,
\KS and $\pi^0$ reconstruction, and charged-kaon identification. In
addition, there are additional contributions to these uncertainties
originating from the disagreement between data and MC distributions
for all the variables used in the selection. These are estimated by
comparing the data and simulation performance in control samples. To
evaluate the uncertainties arising from peaking background
contributions, we repeat the fit by varying the numbers of these
events within their statistical errors. The uncertainties on the
branching fractions of the sub-decay modes are also taken into
account. The uncertainty on $N_{\BB}$ ($1.1\%$) has a negligible
effect on the total error.

The systematic uncertainties on the branching fractions are
summarized in Table \ref{tab:syst}. All the uncertainties are
considered to be uncorrelated and are treated separately for each
channel.

\begin{table*}[htb]
\renewcommand{\arraystretch}{1.3}
\begin{center}
\caption{ Systematic errors on branching fractions for $B^+ \to \Dp
K^0$ and $B^+ \to \Dp \Kstarz$ decay channels. All quantities are
given in units of $10^{-6}$. \label{tab:syst}}
\begin{tabular}{lcccc@{\hspace{8mm}}cc}
\hline\hline
 & \multicolumn{4}{c}{$B^+\to \Dp K^0$ } & \multicolumn{2}{c}{$B^+\to \Dp K^{*0}$ } \\ \cline{2-7}
 & $K\pi\pi $ & $ {K\pi\pi\pi^0}$ & ${\KS\pi}$ & ${\KS\pi\pi^0}$ & $K\pi\pi $ & ${\KS\pi}$ \\
\hline
~~PDF - MC & $^{+0.8}_{-0.8}$ & $^{+6.2}_{-3.4}$ & $^{+5.3}_{-4.4}$ & $^{+7.3}_{-8.8}$ & $^{+0.6}_{-0.9}$ & $^{+3.1}_{-3.6}$ \\
\hline
\multicolumn{7}{l}{Data-MC PDF shapes:}\\
\hline
~~Continuum background &0.2& 0.4& 1.4& 0.5 & 0.1 & 1.7\\
~~\BB\ background &0.7& 1.6& 2.5& 5.0 & 1.0 & 4.4\\
~~Signal &$<0.05$& 9.2& 5.6& 0.9 & 0.9 & 3.1\\
\hline \multicolumn{7}{l}{Efficiency error:}\\ \hline
~~Reconstruction efficiency (MC)&0.1& 0.6&$<0.05$& 0.9 & 0.1 & 0.5\\
~~Data-MC &0.2& 0.8&$<0.05$& 0.5 & 0.2 & 0.3\\
\hline
Peaking background &$<0.05$& 0.5& 0.2& 0.2 &$<0.05$ & 0.1\\
\BR\ errors &0.3& 0.3&$<0.05$& 0.4 &$<0.05$ & 0.1\\\hline \hline
Combined &$^{+1.1}_{-1.3}$ & $^{+11.3}_{ -11.8}$ & $^{+8.2 }_{ -9.3 }$ & $^{+9.0 }_{ -12.5 }$ & $^{+1.5 }_{ -1.8}$ & $^{+6.4 }_{ -7.4}$ \\
\hline\hline
\end{tabular}
\end{center}
\end{table*}

\section{Results for Branching Fractions}

The final likelihood for each decay mode is obtained by convolving
the likelihoods for the measured branching fractions with Gaussian
functions of width equal to the systematic uncertainty.

The final results including systematic uncertainties are
\begin{eqnarray*}
\BR (B^+ \to \Dp K^0)&=& (-3.8\,^{+2.5}_{-2.4})\times 10^{-6},\\
\BR (B^+ \to \Dp K^{*0})&=& (-5.3\pm 2.7)\times 10^{-6}.
\end{eqnarray*}

Since the measurements for the branching fractions are not
statistically significant, following a Bayesian approach and
assuming a flat prior distribution for the branching fractions, we
integrate over the positive portion of the likelihood function to
obtain the following upper limits at 90\% probability:
\begin{eqnarray*}
\BR(B^+ \to \Dp K^0) &<& 2.9\times 10^{-6},\\
\BR(B^+ \to \Dp K^{*0}) &<& 3.0\times 10^{-6}.
\end{eqnarray*}
The $\Bp\to\Dp\Kz$ result represents an improvement over, and
supersedes, our previous result~\cite{cite:polci}, while the
$\Bp\to\Dp\Kstarz$ result is the first for this channel.

\section{Conclusions}

In summary, we have presented a search for the rare decays
$\Bp\to\Dp\Kz$ and $\Bp\to\Dp\Kstarz$, which are predicted to
proceed through annihilation or rescattering amplitudes. We do not
observe any significant signal and we set 90\% probability upper
limits on their branching fractions.

\section{Acknowledgments}

We are grateful for the 
extraordinary contributions of our \pep2\ colleagues in
achieving the excellent luminosity and machine conditions
that have made this work possible.
The success of this project also relies critically on the 
expertise and dedication of the computing organizations that 
support \babar.
The collaborating institutions wish to thank 
SLAC for its support and the kind hospitality extended to them. 
This work is supported by the
US Department of Energy
and National Science Foundation, the
Natural Sciences and Engineering Research Council (Canada),
the Commissariat \`a l'Energie Atomique and
Institut National de Physique Nucl\'eaire et de Physique des Particules
(France), the
Bundesministerium f\"ur Bildung und Forschung and
Deutsche Forschungsgemeinschaft
(Germany), the
Istituto Nazionale di Fisica Nucleare (Italy),
the Foundation for Fundamental Research on Matter (The Netherlands),
the Research Council of Norway, the
Ministry of Education and Science of the Russian Federation, 
Ministerio de Ciencia e Innovaci\'on (Spain), and the
Science and Technology Facilities Council (United Kingdom).
Individuals have received support from 
the Marie-Curie IEF program (European Union), the A. P. Sloan Foundation (USA) 
and the Binational Science Foundation (USA-Israel).


\begin{thebibliography}{99}
\bibitem{cite:ch} Charge conjugation is implied throughout this paper.
\end{thebibliography}
\end{document}